\documentclass[twocolumn]{aastex62}
\usepackage{natbib}
\usepackage{color,ulem}
\graphicspath{{./}{../pictures/}}

\received{}
\revised{}
\accepted{}
\submitjournal{ApJ}

\shorttitle{Tidal disruption events}
\shortauthors{Ryu et al.}

\usepackage{graphicx}	
\usepackage{amsmath}	
\usepackage{amssymb}	
\usepackage{cases}
\usepackage{array,multirow}
\usepackage{upgreek}
\usepackage{textcomp}

\usepackage{hyperref}
\usepackage[referable]{threeparttablex}
\usepackage{booktabs, dcolumn}
\makeatletter
\newcommand*{\rom}[1]{\expandafter\@slowromancap\romannumeral #1@}
\makeatother

\newcommand{\beq}{\begin{equation}}
\newcommand{\eeq}{\end{equation}}
\newcommand{\simlt}{\mathrel{\hbox{\rlap{\hbox{\lower4pt\hbox{$\sim$}}}\hbox{$<$}}}}
\newcommand{\simgt}{\mathrel{\hbox{\rlap{\hbox{\lower4pt\hbox{$\sim$}}}\hbox{$>$}}}}

\newcommand{\Msol}{\;M_{\odot}}
\newcommand{\Rsol}{\;R_{\odot}}

\newcommand{\Mbh}{M_{\bullet}}
\newcommand{\Mstar}{M_{\star}}
\newcommand{\Rstar}{R_{\star}}

\newcommand{\harm}{{\sc Harm3d}}   
\newcommand{\mesa}{{\small MESA}}
\newcommand{\patchwork}{{\sc Patchwork}}
\newcommand{\patchworkmhd}{{\sc PatchworkMHD}}

\def\apjl{ApJL}
\def\apj{ApJ}
\def\mnras{M.N.R.A.S.}
\def\aap{A\&A}
\def\nat{Nat.}

\def\pasp{PASP}
\def\apjs{ApJ Supp.}
\def\aj{AJ}

\usepackage[normalem]{ulem}

\defcitealias{Ryu1+2019}{Paper 1}
\defcitealias{Ryu2+2019}{Paper 2}
\defcitealias{Ryu3+2019}{Paper 3}
\defcitealias{Ryu4+2019}{Paper 4}

\begin{document}

\title{Shocks Power Tidal Disruption Events}

\correspondingauthor{Taeho Ryu}
\email{tryu@mpa-garching.mpg.de}

\author[0000-0002-0786-7307]{Taeho Ryu}
\affil{Max Planck Institute for Astrophysics, Karl-Schwarzschild-Strasse 1, 85748 Garching, Germany}
\affil{Physics and Astronomy Department, Johns Hopkins University, Baltimore, MD 21218, USA}

\author{Julian Krolik}
\affiliation{Physics and Astronomy Department, Johns Hopkins University, Baltimore, MD 21218, USA}

\author{Tsvi Piran}
\affiliation{Racah Institute of Physics, Hebrew University, Jerusalem 91904, Israel}

\author{Scott C. Noble}
\affiliation{Gravitational Astrophysics Laboratory, NASA Goddard Space Flight Center, Greenbelt, MD 20771, USA}

\author{Mark Avara}
\affiliation{Institute of Astronomy, University of Cambridge, Madingley Road, Cambridgeshire, England}

\begin{abstract}

Accretion of debris seems to be the natural mechanism to power the radiation emitted during a tidal disruption event (TDE), in which a supermassive black hole tears apart a star. However, this requires the prompt formation of a compact accretion disk. Here, using a fully relativistic global simulation for the long-term evolution of debris in a TDE with realistic initial conditions, we show that at most a tiny fraction of the bound mass enters such a disk on the timescale of observed flares.  To ``circularize" most of the bound mass entails an increase in the binding energy of that mass by a factor $\sim 30$; we find at most an order unity change.  Our simulation suggests it would take a time scale comparable to a few tens of the characteristic mass fallback time to dissipate enough energy for ``circularization".  Instead, the bound debris forms an extended eccentric accretion flow with eccentricity $\simeq 0.4-0.5$ by $\sim 2$ fallback times.  Although the energy dissipated in shocks in this large-scale flow is much smaller than the ``circularization" energy, it matches the observed radiated energy very well. Nonetheless, the impact of shocks is not strong enough to unbind initially bound debris into an outflow.

\end{abstract}


\section{Introduction}

A tidal disruption event (TDE) takes place when a star that has wondered into the vicinity of a supermassive black hole (SMBH) is torn apart by the SMBH's gravitational field.  About half of the stellar mass is left unbound and is ejected to infinity while the other (bound) half returns to the vicinity of the black hole. The end result is a flare of optical/UV, X-ray, and at times, radio emission.  

During the last decades, TDEs were transformed from a theoretical prediction \citep{Hills1988,Rees1988} to an observational reality \citep{Gezari2021}.  With the use of numerous detectors, ranging from ROSAT (and now eROSITA) in X-rays, to GALEX in the UV, and multiple telescopes in the optical band (including systematic surveys like SDSS \citep{SDSS}, \citep{SDSS}, ASSAS-N \citep{ASASSN},
Pan-STARRS \citep{PANSTARRS},
and, more recently, ZTF \citep{ZTF},
more than a hundred TDEs have now been detected.   Upcoming observations with the Rubin Observatory will provide an overwhelming amount of data in the near future. 

Although TDEs are of great interest in their own right, their properties offer a wide range of opportunities to learn about other astrophysical questions.  They can reveal quiescent SMBHs and possibly permit inference of their masses. 
They present otherwise unobtainable information about non-steady accretion onto SMBHs and the conditions for jet-launching. In addition, understanding the rates of TDEs  would reveal valuable information concerning the stellar dynamics in galaxy cores. 

Early theoretical predictions pointed out that  the  bound  debris returns to the BH at a rate $\propto t^{-5/3}$ \citep{Rees1988,Phinney1989}.  It was then speculated that, having returned, matter quickly forms a disk whose outer radius is comparable in size to the pericenter of the star's original trajectory.  With such a small scale, the inflow time through the disk should be so short that the light emitted by the disk would follow the matter fallback, i.e., likewise $\propto t^{-5/3}$.   With a characteristic temperature $\sim 10^5 - 10^6$~K, the emitted light would be in the FUV/EUV or soft X-ray band, and the peak luminosity would be much larger than Eddington. 

However,  the observed luminosity rarely reaches the Eddington luminosity for the expected SMBH masses. In addition, when more TDEs were discovered in the optical, it was realized that typical temperatures are a few $10^4$ K \citep{Gezari2021}, implying that the radiating area is much larger than that of a small disk whose radial scale is similar to the star's pericenter.  Moreover, the total energy radiated during the flare period is generally two orders of magnitude smaller than the energy expected if half the stellar mass had been efficiently accreted onto a SMBH.  In fact, it is an order of magnitude smaller than the ``circularization" energy that would have been emitted during the initial formation of the small disk envisioned. That the radiated energy is so small is sometimes called the ``inverse energy crisis" \citep{Piran+2015,Svirski+2017}.

A possible explanation for the low luminosity and low temperatures observed is that the energy produced by the accretion disk is reprocessed by a radiation-driven wind ejected from the disk itself \citep{Strubbe+2009}. If this wind carries a significant kinetic energy, it would also resolve the ``inverse energy crisis" \citep{MetzgerStone2016} (see, however \citealt{MatsumotoPiran2021}). An alternative possibility is that matter does not circularize quickly and most of the observed emission arises from self-intersection shocks at the apocenter \citep{Shiokawa+2015,Piran+2015, Krolik+2016}. These shocks, which are expected to be strong at and shortly after the time of maximum mass return, take place at $\simeq O(10^{3})r_{\rm g}$ from the BH ($r_{\rm g}\equiv G\Mbh/c^{2}$ for BH mass $\Mbh$), are consistent with the observed luminosity, temperature, line width and total energy generated \citep{TDEmass}.

The long-term fate of bound debris is less clear.   Because it is created on highly eccentric orbits, with only a small further diminution in angular momentum it may be able to fall directly into the black hole, releasing very little energy \citep{Svirski+2017}. 
 Alternatively, when there has been time for the magnetorotational instability (MRI) to build strong magnetohydrodynamic (MHD) turbulence and for the gas to lose energy to radiation, the debris may accrete gradually, while radiating efficiently \citep{Shiokawa+2015}.

Although it may be possible by observational means to determine whether one of these two different scenarios occurs, numerical simulations of the disruption and subsequent accretion process may provide an alternative way to resolve this issue. However, such simulations are hindered by the extreme contrast in length and timescales involved. Adding fully relativistic features that are critical for some of the physics ingredients also poses technical challenges.

Aiming to clarify the many questions regarding the evolution of the bound debris' energy, angular momentum, and location,
in this work we present a fully relativistic numerical simulation of a complete tidal disruption of a realistic $3M_\odot$ star by a $10^5 M_\odot$ SMBH in which we follow the system long enough to see the majority of the bound mass return to the black hole. Several of these features have never previously appeared in a global TDE simulation: fully relativistic treatment, both in hydrodynamics and stellar self-gravity; a main-sequence internal structure for the star's initial state; and its long duration.
Our simulation scheme does not include time-dependent radiation transfer; instead, we make the approximation (for our parameters, well-justified for most of the debris mass) that the radiation pressure is the value achieved in LTE and does not diffuse relative to the gas.

The structure of the work is as follows: we define the physical problem and present our numerical scheme in  \S   \ref{sec:methodology}. We discuss the results in \S \ref{sec:results} and their implications in \S \ref{sec:discussion}. In this last section we also compare our results to previous work.   We conclude in \S \ref{sec:Summary}, where we also discuss the possible observational implications of this work. 

\section{The Calculation}\label{sec:methodology}

\subsection{Physics Problem}\label{sec:model}
To provide a context for our choice of numerical methods and parameters, we begin with a brief discussion of our problem's overall structure.  
We begin with a $3\Msol$ star (radius $\Rstar= 2.4\Rsol$) whose internal structure is taken from a \mesa ~\citep{Paxton+2011} evolution to middle-age on the main-sequence, the age at which the hydrogen mass fraction in its core has fallen to 0.5.   The star (initially placed at $r \simeq 900~r_{\rm g}$ from the black hole) approaches a SMBH of  $\Mbh=10^5 \Msol$ on a parabolic orbit with a pericenter distance of $r_{\rm p}\simeq 110~r_{\rm g}$, just close enough to be fully disrupted. Although the nominal tidal radius $r_{\rm t} = \Rstar (\Mbh/\Mstar)^{1/3}$ is $\simeq 370~r_{\rm g}$ for this black hole mass and stellar mass, the critical distance within which a star is fully disrupted is given by $\Psi(\Mbh, \Mstar)r_{\rm t}$, where the order-unity factor $\Psi$ encodes general relativistic corrections through its $\Mbh$ dependence and stellar structure corrections through its $\Mstar$ dependence \cite{Ryu+2020a}.

As the star passes through pericenter, it begins to be torn apart; the process is complete by the time its center-of-mass reaches $\sim7500 r_{\rm g}$.
 During the debris' first orbit, its trajectory is ballistic, with specific orbital energy $E$ and specific angular momentum $J$ very close to that of the star's center-of-mass.  Immediately after the disruption, the distribution of mass with energy $dM/dE$ is roughly a square-wave and is approximately symmetric with respect to $E=0$.  Although order-of-magnitude arguments suggest that the half-width of the energy square-wave $\Delta\varepsilon\simeq G\Mbh \Rstar/r_{\rm t}^2$, \citet{Ryu+2020a} showed that this estimate should be multiplied by an order-unity correction factor $\Xi(\Mbh,\Mstar)$, which depends on both $\Mbh$ and $\Mstar$ as it accounts for both relativistic and stellar internal structure effects.  For our parameters, $\Xi \approx 1.64$, so that the energy half-width is
\begin{align}
    \Delta E 
\simeq 1.4\times 10^{-4}   c^2 &\left(\frac{\Xi}{1.64}\right)\left(\frac{M_{\bullet}}{10^{5}\Msol}\right)^{1/3}\nonumber\\
&\times\left(\frac{M_{\star}}{3\Msol}\right)^{2/3}\left(\frac{R_{\star}}{2.4R_{\odot}}\right)^{-1}\ .
\end{align}
The semimajor axis $a$ and the eccentricity $e$ of the bound debris with $E=-\Delta E$ are
\begin{align}\label{eq:amin}
      a &= \frac{G\Mbh}{2E} \simeq 3600  r_{\rm g}\left(\frac{\Xi} {1.64}\right)^{-1}
      \nonumber\\
      &\left(\frac{\Mbh}{10^{5}\Msol}\right)^{-1/3}\left(\frac{\Mstar}{1\Msol}\right)^{-2/3}\left(\frac{\Rstar}{2.4\Rsol}\right),
  \end{align} 
and
\begin{align}\label{eq:ecc}
   e \simeq 1 - 0.07\left(\frac{\Xi}{1.64}\right)\left(\frac{\Mbh}{10^{5}\Msol}\right)^{-1/3}\left(\frac{\Mstar}{3\Msol}\right)^{1/3},
  \end{align}
respectively. 
The apocenter distance of the debris with $E=-\Delta E$ is then $(1+e)a\simeq 7000  r_{\rm g}$.

The bound debris, i.e., debris with $E<0$, must return to the vicinity of the SMBH. The maximal mass return rate occurs when the debris with $E\simeq -\Delta E$ returns. The fallback time scale of the material with $E\simeq -\Delta E$ is very nearly its orbital period, 
\begin{align}
\label{eq:peak_t}
t_{0}&=\frac{\uppi}{\sqrt{2}}\frac{G M_{\bullet}}
{\Delta {E}^{3/2}}
\simeq 7.6~{\rm days}~\left(\frac{\Xi}{1.64}\right)^{-3/2}\nonumber\\
&\left(\frac{\Mbh}{10^{5}\Msol}\right)^{1/2} \left(\frac{M_{\star}}{3\Msol}\right)^{-1}\left(\frac{R_{\star}}{2.4\Rsol}\right)^{3/2}. 
\end{align}
The maximal fallback rate is then 
\begin{align}
\label{eq:peak_mdot}
\dot{M}_{0}&\simeq \frac{M_{\star}}{3t_{\rm 0}}
\simeq 0.13 \Msol \, {\rm days}^{-1} \left(\frac{\Xi}{1.64}\right)^{3/2}\nonumber\\
&\times\left(\frac{\Mbh}{10^{5}\Msol}\right)^{-1/2}
\left(\frac{M_{\star}}{3\Msol}\right)^{2}\left(\frac{R_{\star}}{2.4\Rsol}\right)^{-3/2}.
\end{align}

The different scales involved are why the problem is a difficult one for numerical simulation.  Fluid travels through regions where the characteristic scale on which gravity changes runs from $\sim  r_{\rm g}$ to $\sim 10^4 r_{\rm g}$, while the structure of the star varies on a scale that is a fraction of $\Rstar \sim 15 r_{\rm g}$. Similarly, the characteristic dynamical timescale ranges from $\sim  r_{\rm g}/c$ to $\sim 10^6 r_{\rm g}/c$.

\begin{figure*}
\includegraphics[width=18cm]{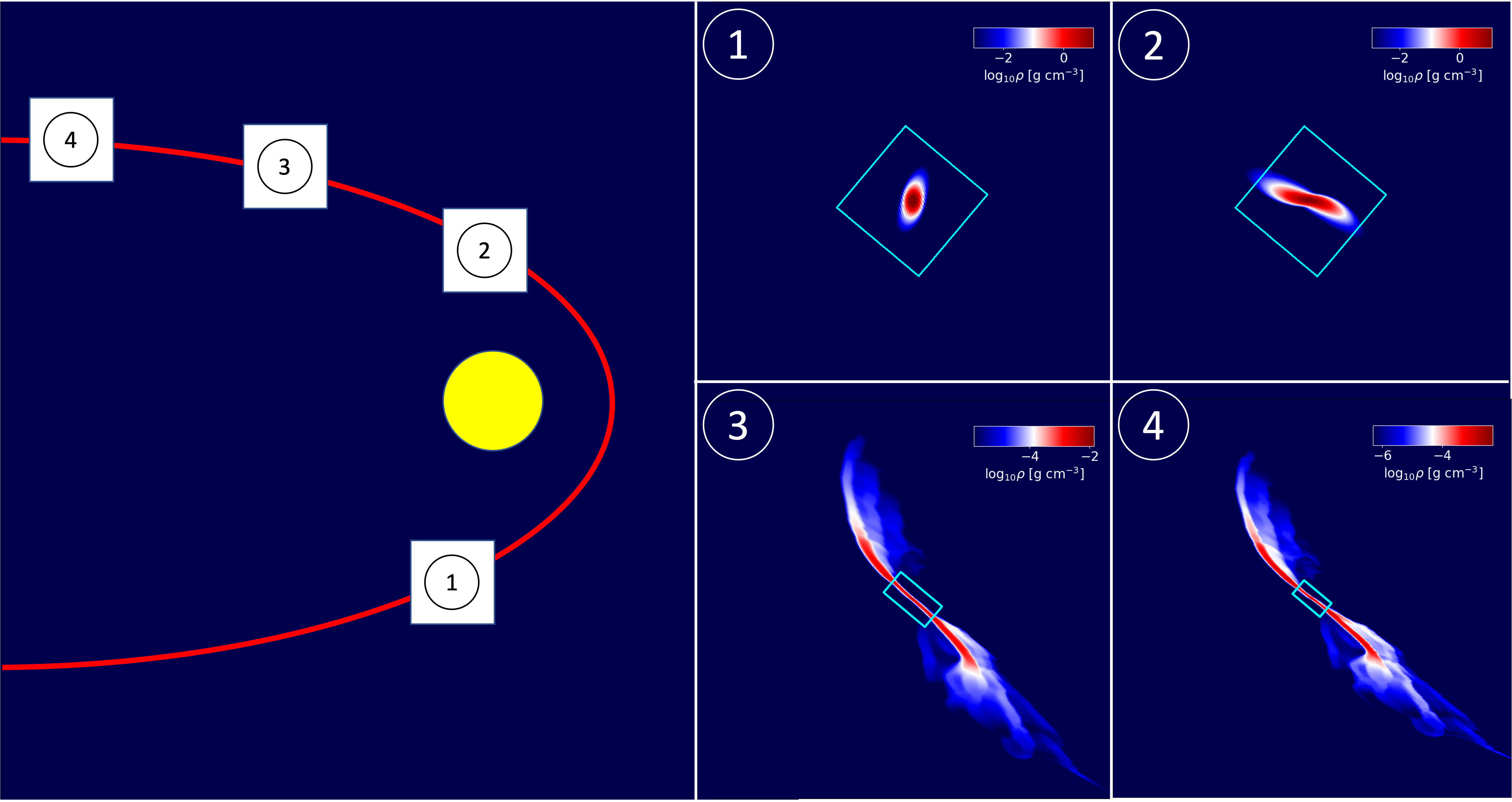}
	\caption{Successive moments in a full tidal disruption event simulated using two domains: The density distribution of the star before (1) and after (2) the pericenter passage and the debris after the pericenter passage (3, 4) in the equatorial plane. The yellow dot in the big \textit{left} panel indicates the BH. The cyan box in the four \textit{right} panels depicts the boundary of \textit{Domain1}. Once the star is disrupted, we increase the size of the box (from 2 to 3) to keep the debris inside \textit{Domain1} where the self-gravity is calculated as much as possible. Panels are not drawn to scale.}
	\label{fig:schematic}
\end{figure*}

\subsection{Code}\label{subsec:code}

To treat this problem, we perform a fully relativistic global hydrodynamics simulation using a software package comprising three core codes, \patchworkmhd~ (Avara et al., in prepartion), \harm~\citep{Noble+2009}, and a relativistic self-gravity solver \citep{Ryu+2020b}.

We use the intrinsically conservative general relativistic magneto-hydrodynamics (GRMHD) code \harm \ \citep{Noble+2009} to solve the equations of relativistic pure hydrodynamics.  This code employs the Lax–Friedrichs numerical flux formula and uses a parabolic interpolation method \citep{ColellaWoodward1984} with a monotonized central-differenced slope limiter. Because of its robust algorithm, it has been used for studying a wide variety of SMBH accretion problems \citep[e.g.,][]{Noble+2009,Noble+2012,Shiokawa+2015, Ryu+2020a, Ryu+2020b}. We take radiation pressure into account by setting the pressure $p = \rho kT/\bar{m} + aT^4/3$ when the internal energy  density $u = (3/2)\rho kT/\bar{m} + aT^4$.  Here $\rho$ is the proper rest-mass density, $\bar{m}$ is the mass per particle, and $T$ is the temperature.  We can then define an equation of state with an ``effective adiabatic index'' \citep{Shiokawa+2015} that varies between $\gamma=4/3$ and $5/3$  depending on the ratio of the gas pressure and radiation pressure. 

Our self-gravity solver is described in detail in \cite{Ryu+2020b}.  In brief, it constructs a metric valid in the star's center-of-mass frame by superposing the potential found by a Poisson solver operating in a tetrad frame defined at the center-of-mass on top of the metric for the BH's Kerr spacetime described in the center-of-mass frame.   Because this is a compact free-fall frame, the spacetime for the star and its close surroundings is always very nearly flat.

The large contrasts in length and time scales noted earlier make such a computation prohibitively expensive if the entire range of scales is resolved in a single domain. To overcome this difficulty, we introduce multiple domains with \patchworkmhd (Avara et al., in preparation, first application in \citealt{Avara+2023}). \patchworkmhd~is an extension of the original purely hydrodynamic \patchwork~code \citep{Shiokawa+2018} that both enables MHD and introduces a number of algorithmic optimizations. 
Both versions create a multiframe / multiphysics / multiscale infrastructure in which independent programs simultaneously evolve individual patches with their own velocities, internal coordinate systems, grids, and physics repertories.  The evolution of each patch is parallelized in terms of subdomains; where different patches overlap, the infrastructure coordinates boundary data exchange between the relevant processors in the different patches.  A further extension of PatchworkMHD, specialized to problems involving numerical relativity, is described in  \citet{patchworkwave}.

\subsection{Domain setup}

During the first part of a TDE, a star travels through extremely  rarefied gas as it traverses a nearly parabolic trajectory around the black hole.  This situation is a prime example of the contrasts motivating our use of a multipatch system: the characteristic lengthscales inside the star are much  smaller than those in  the surrounding gas; in addition, self-gravity is important inside and near the star, but not in the remainder of the volume. Consequently, during this portion of the event we employ \patchworkmhd~and apply it to two patches: one covering the star, the other covering the remainder of the SMBH's neighborhood.  Once the star is fully disrupted, there is no further need of the star patch, and it is removed, its content interpolated onto a single remaining patch, which evolves the entire region around the SMBH.

\subsubsection{The star's pericenter passage---two-patch simulation}

This  first stage begins with the initial approach of the star to pericenter and ends when the star's center-of-mass trajectory reaches a distance from the SMBH $\simeq 20r_{\rm t}$.  It ends here because the star is then fully disrupted.  During this stage, a rectangular-solid Cartesian domain denoted \textit{Domain1} covers the volume around the star; it is completely embedded in a larger spherical-coordinate domain denoted \textit{Domain2} that ultimately covers a large spherical region centered on the SMBH.

Following the methods described in \citet{Ryu+2020b}, in {\it Domain1} we follow the hydrodynamics of gas with self-gravity in a frame that follows the star's center of mass. Initially, the domain is a cube with edge-length $5R_{\star}$, and the cell size in each dimension is $\simeq R_{\star}/25$.  The orientation of this box relative to the black hole is rotated during the disruption in order to follow the direction of the bulk of the tidal flow, which is easily predicted from  Kepler's laws.  As the debris expands, the cube is adaptively extended with constant cell-size to keep the debris inside the domain for as long as possible. 

As the debris expands,  a fraction of the debris crosses smoothly from \textit{Domain1} to \textit{Domain2}, where we continue to evolve it under the gravity of the SMBH without any gas self-gravity. For \textit{Domain2}, we adopt modified spherical coordinates in Schwarzschild spacetime.  That is, if the code's three spatial coordinates are $(x_1,x_2,x_3)$,
they can represent spherical coordinates $(r,\theta,\phi)$ through the relations
\begin{align}\label{eq:coord1}
    r &= e^{x_1},\nonumber\\
     \theta & = 0.5\uppi [ 1.0 + h_{1} x_2 + (1 - h_{1} - \frac{2.0 \theta_{0}}{\uppi}) {x_2}^{h_{2}}],\nonumber\\
    \phi &= x_3.
\end{align}
 where $h_{1}$ and $h_{2}$ are tuning parameters that determine the vertical coordinate structure ($h_{1}\simeq 0.03$ and $h_{2}\simeq 9$), and $\theta_{0}$ is the angle from the polar axis to the $\theta$-boundaries.  In this coordinate system, the radial grid cells have a constant ratio of cell-dimension to radius, and the $\theta$ grid cells are concentrated near the mid-plane. This coordinate system is suitable for modeling systems that involve a wide range of radial scales and contain a disk-like structure near the mid-plane.
 
 To be computationally efficient, instead of fixing the size and the tuning parameters throughout the simulation, we flexibly adjust the size and the resolution of \textit{Domain2} so that it is large enough to contain the entire debris, but we do not waste cells on regions where there is no debris.   This strategy reduces the computational cost by a large factor.  To ensure proper resolution during the period of flexible domain size, we require at least 15-20 cells per scale height in all three dimensions.  In addition, at all times during the two-patch evolution, we keep the cell sizes in the overlapping regions of both domains  comparable. 

At its largest extent, {\it Domain2} runs from $r_{\rm min} = 40r_{\rm g}$ to $r_{\rm max} = 18000r_{\rm g}$.  The maximum radius is chosen to be greater than the apocenter of debris that would return to the black hole within a time $4t_0$; our simulation ran for $3t_0$.   The minimum radius was chosen by balancing two opposing goals: minimizing the mass lost through the inner radial boundary while maximizing the time step so as to limit computational expense. Similarly, when {\it Domain2} has its largest volume, the polar angle extends from $\theta_0 = 2^\circ$ to $\pi - \theta_0 = 178^\circ$.  The azimuthal angle $\phi$ covers a full $2\pi$ when {\it Domain2} is largest.

We adopt outflow boundary conditions at all boundaries of \textit{Domain2}. All the primitive variables are extrapolated to the ghost cells with zero gradients. However, to ensure outflow, if the extrapolated normal component of the fluid velocity in the ghost cells is directed inward, it is set to zero.
When this domain is maximally  extended azimuthally, we provide boundary conditions to the processor domains having surfaces at $\phi = 0$ and $\phi = 2\pi$ by matching those with the same radial and polar angle locations.

We show in Figure~\ref{fig:schematic} the density distribution of the star before the pericenter passage (\textit{top-left} of the four small panels), during the passage (\textit{top-right}) and the debris after the passage (\textit{bottom}) in the equatorial plane in the two-domain simulation. The cyan box demarcates the boundary of \textit{Domain1}. These figures demonstrate how, as the star is  disrupted, the stellar debris crosses smoothly the inter-patch boundary between \textit{Domain1} and \textit{Domain2}.

\subsubsection{Single-domain simulation}

When only a tiny fraction of the star's original mass remains within {\it Domain1}, self-gravity becomes irrelevant (measured in the star center-of-mass frame, the acceleration throughout {\it Domain 1} due to self-gravity is $\lesssim 10\%$ of the tidal acceleration), and gradients in gas properties are no longer connected with $\Rstar$.  We therefore remove {\it Domain1} and continue the simulation using only {\it Domain2}.  In this stage, it takes its maximum extent.

Shocks caused by the stream-stream intersection dissipate the orbital energy into thermal energy, which results in the vertical expansion of streams. The vertical coordinate system used for the early evolution (Equation~\ref{eq:coord1}) is not suitable to resolve gas at a large height as the resolution becomes increasingly crude as $z$ increases. To better resolve the gas at high $z$, we reduce the concentration of $\theta$ cells to the midplane while leaving the $r$ and $\phi$ grids untouched.  To do this, we redefine $\theta(x_2)$:
\begin{align}
    \theta & = \alpha(\tanh[b(x_2 - a)] + \tanh[b(x_2 + a)]) + 0.5\uppi,
\end{align}
where $\alpha= -(0.5\uppi- \theta_{0})/[\tanh(b(-0.5 - a))+\tanh(b(-0.5 + a))]$.
Here, $a$ and $b$ are a set of tuning parameters that determine the vertical structure. At this stage, we fix the domain extent of $r$ and $\phi$. But we keep adjusting the domain extent and the resolution of the vertical structure flexibly, whenever it is necessary, to ensure sufficiently high vertical resolution (at least 20 cells per vertical scale heights) by properly choosing the cell number (within $80-120$), $\theta_{0}$ ($2-15^{\circ}$) and the tuning parameters ($a\simeq 0.32-0.34$ and $b\sim 9.8$).

\begin{figure}
\includegraphics[width=8.6cm]{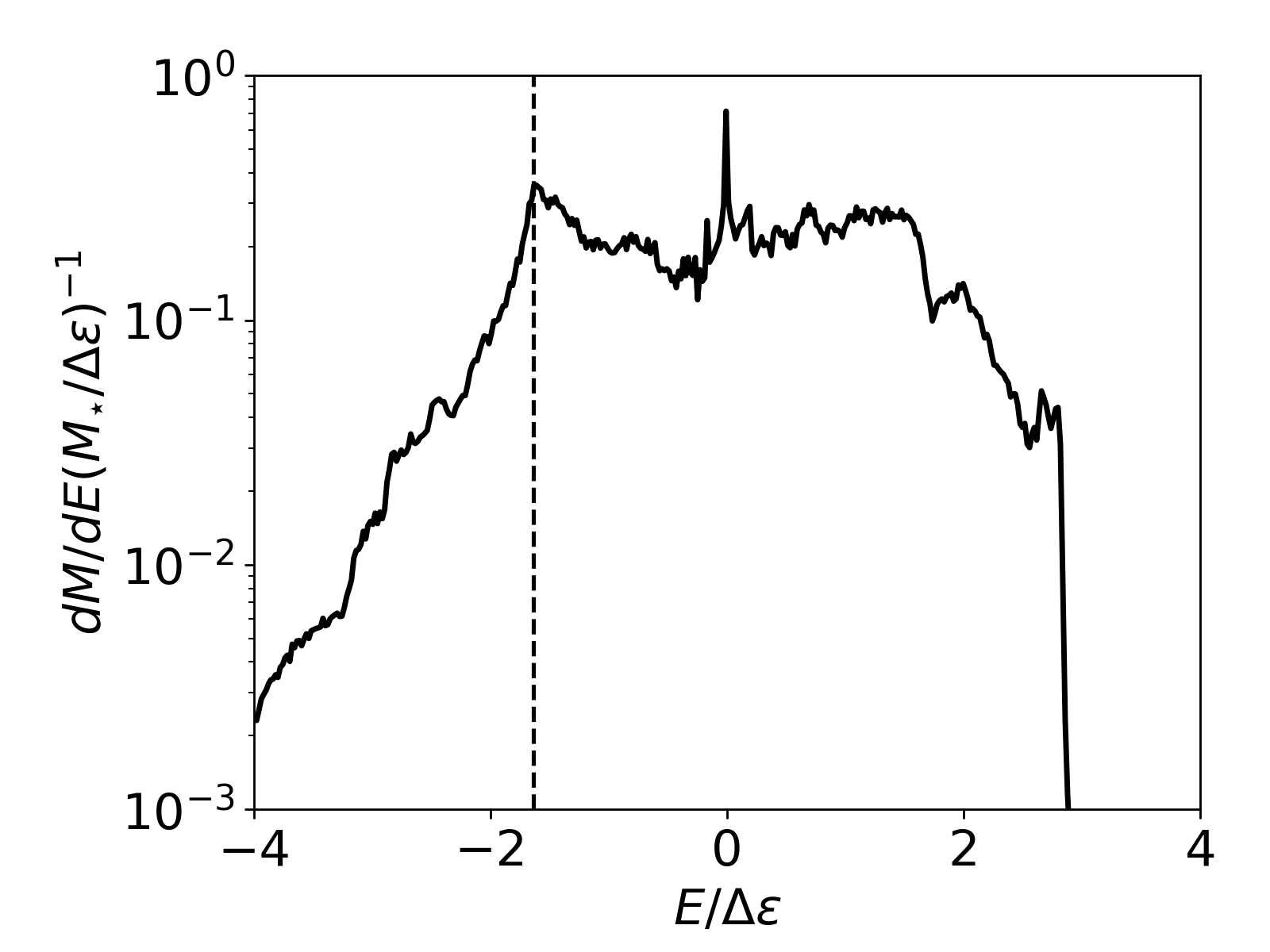}
	\caption{The energy distribution of debris at $t\simeq 0.4t_{0}$. The vertical dashed line ($E/\Delta\epsilon=\Xi\simeq-1.64$) shows the characteristic energy at which fallback rate peaks (see Figure~\ref{fig:massreturnrate}). }
	\label{fig:debris_e}
\end{figure}

\begin{figure}
\includegraphics[width=8.6cm]{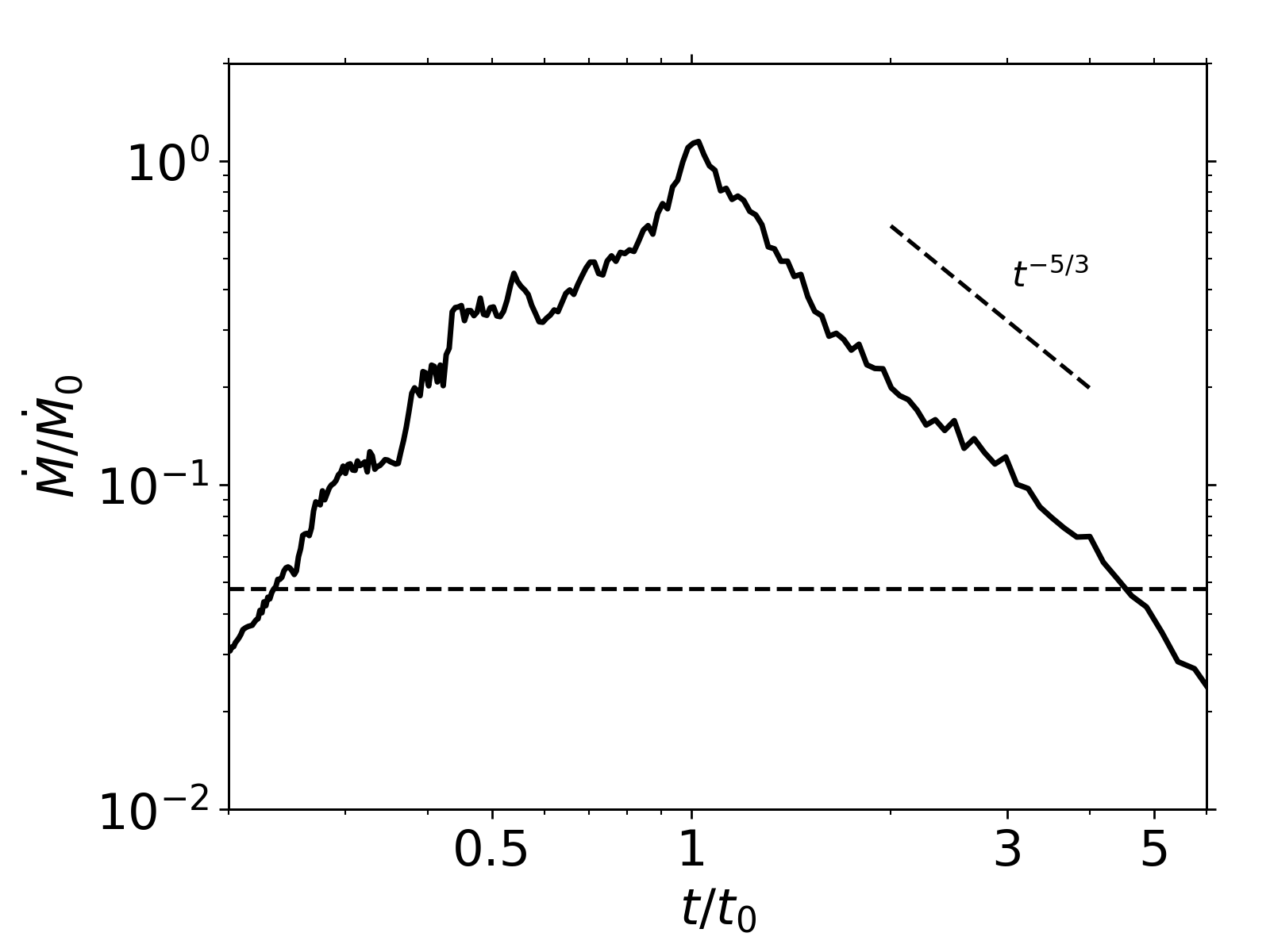} 
	\caption{The mass fallback rate $\dot{M}$, using the energy distribution shown in Figure~\ref{fig:debris_e}. The rate and time are normalized by $\dot{M}_{0}=M_{\star}/3t_{0}$ and $t_{0}$, respectively. The horizontal dashed line indicates the Eddington limit with the radiative efficiencty of 0.01 and the diagonal dashed line a power-law of $t^{-5/3}$. }
	\label{fig:massreturnrate}
\end{figure}

\begin{figure}\centering
\includegraphics[width=5.8cm]{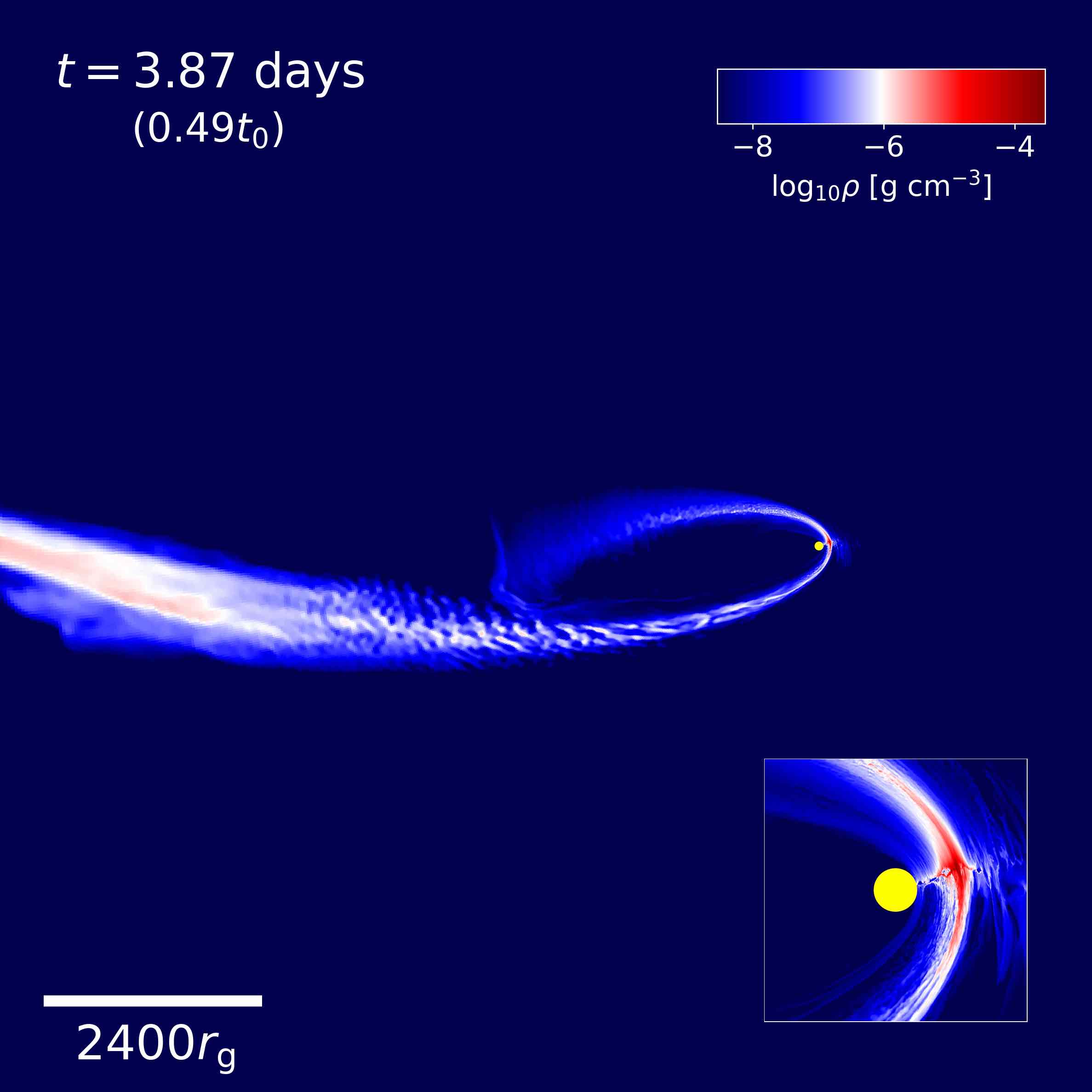}
\includegraphics[width=5.8cm]{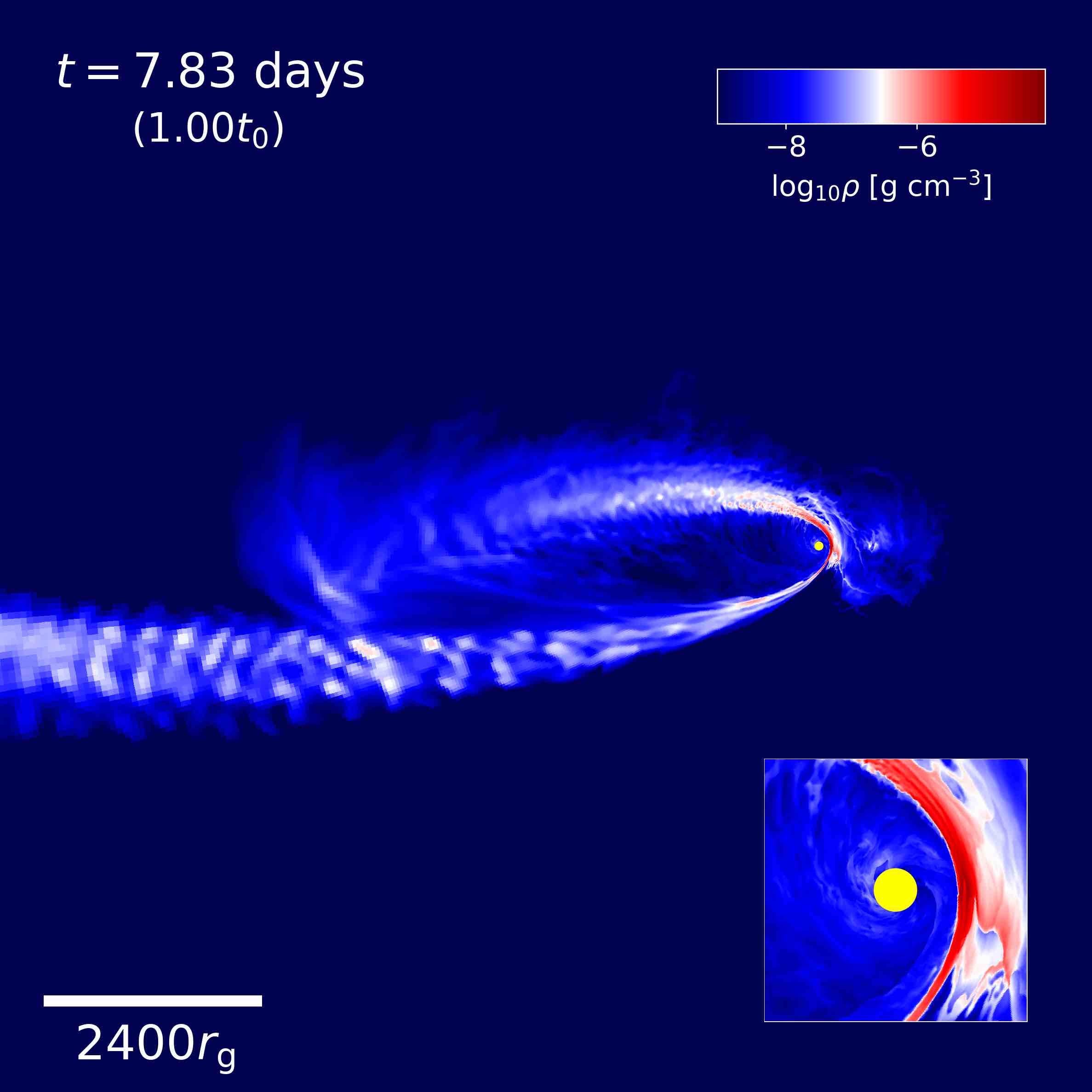}
\includegraphics[width=5.8cm]{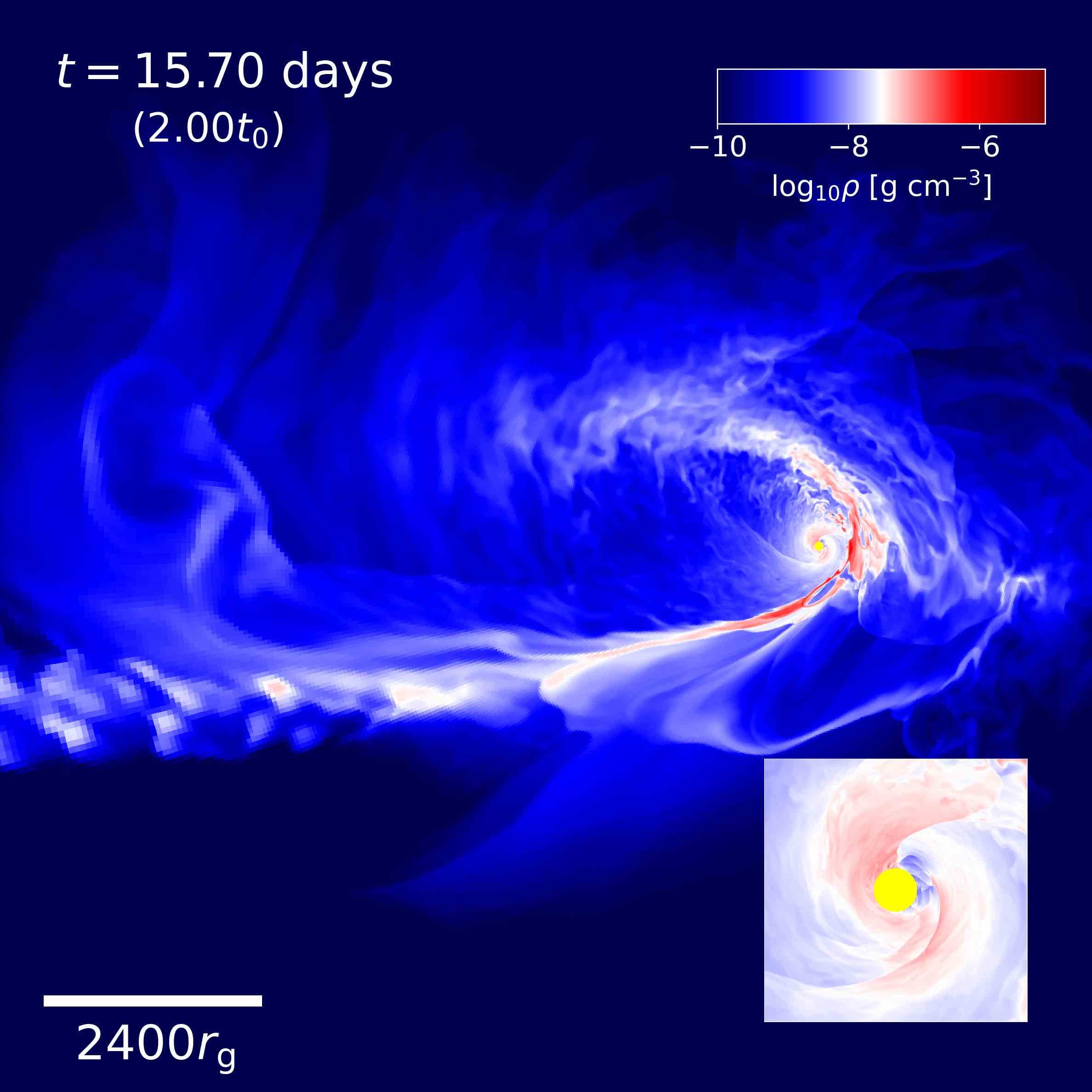}
\includegraphics[width=5.8cm]{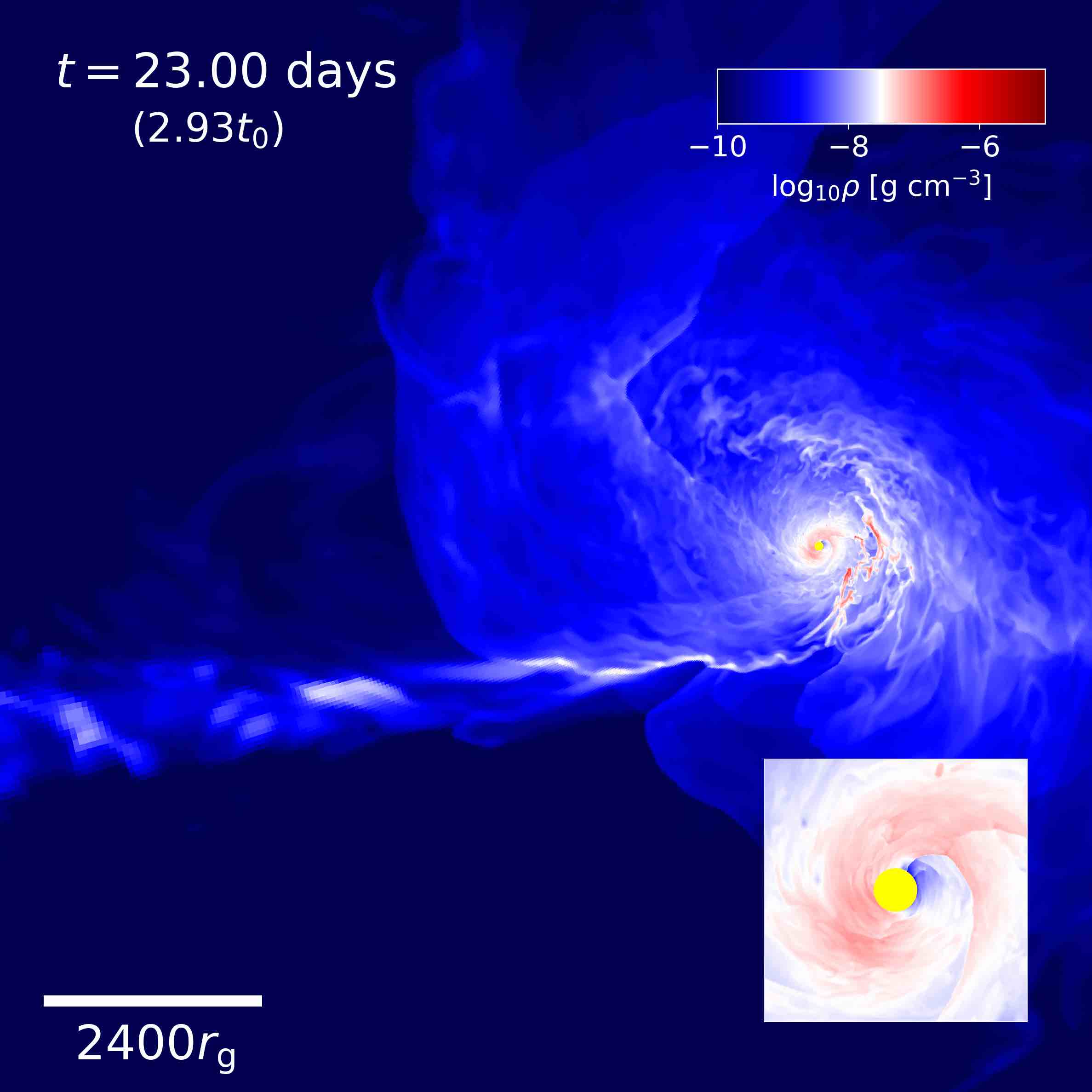}
	\caption{The density distribution around the black hole excision (yellow dot) in the equatorial plane at four different times, $t/t_{0}=0.5$, $1$, $2$ and 3. The extent of the inset is $500r_{\rm g}$.}
	\label{fig:densitydistribution}
\end{figure}

\section{Results}
\label{sec:results}

\begin{figure}
\includegraphics[width=8.6cm]{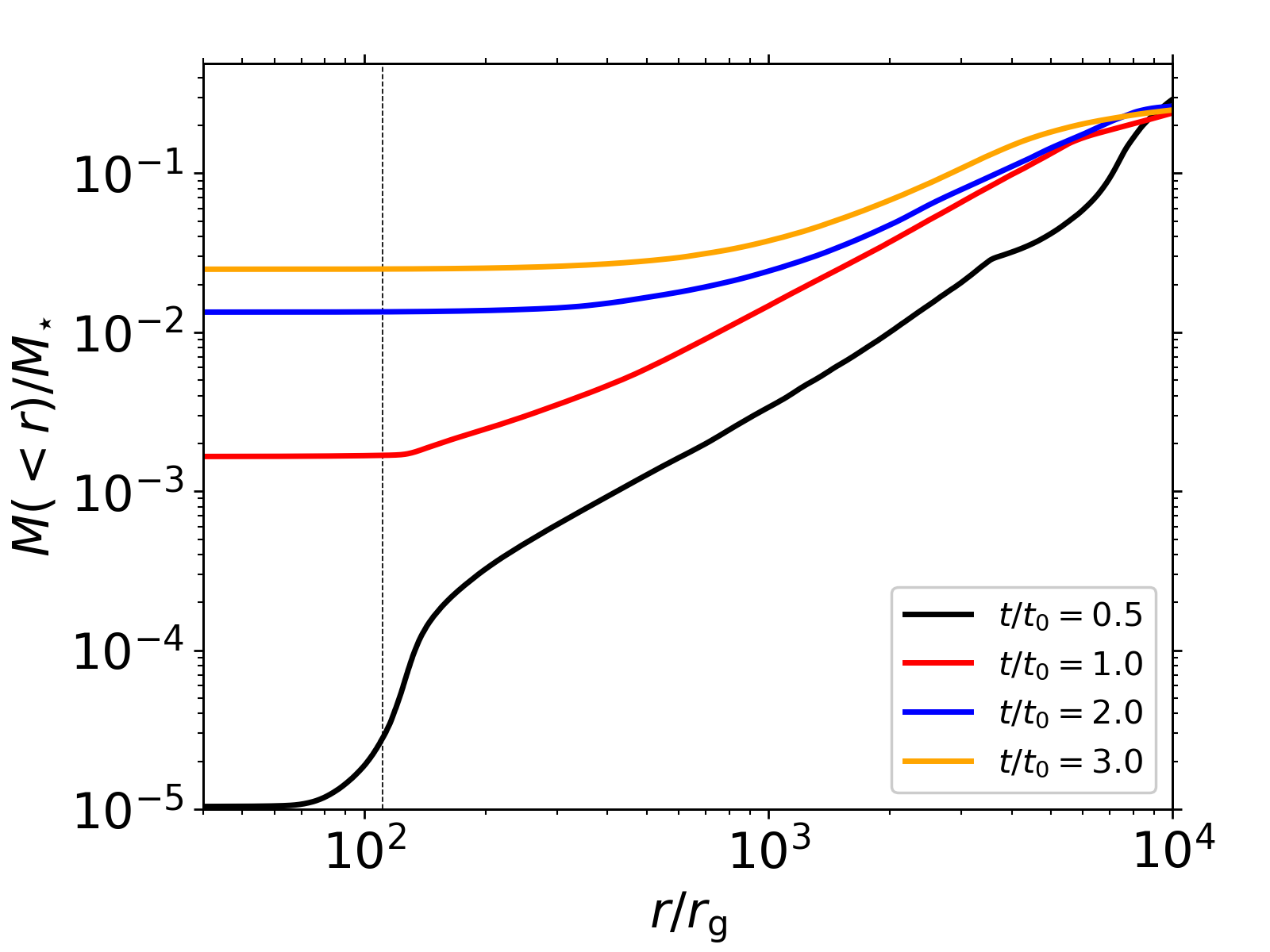}
	\caption{Accumulated mass normalized by $M_{\star}$ 
 as a function of the distance from the SMBH at $t/t_{0}\simeq 0.5$, 1, 2, and 3. The dashed black vertical line indicates the pericenter distance of the original stellar orbit. The mass expelled through the radial inner boundary at $r=40r_{\rm g}$ is included as if the accreted mass is confined within $r=40r_{\rm g}$. }
	\label{fig:massenergybudget}
\end{figure}

\subsection{Overview}
\label{sub:overview}

The star becomes strongly distorted as it passes through the pericenter (see Figure~\ref{fig:schematic}) and then falls apart entirely as it travels farther away from the black hole.  When its entire mass has been dispersed,
the orbital energy distribution of the debris $dM/dE$ is not confined within the order of magnitude estimate of the energy distribution's width $\Delta\epsilon$; it is about twice as wide and has extended wings (Figure~\ref{fig:debris_e}). The mass of bound gas ($E<0$) is $\simeq 0.48M_{\star}$, slightly smaller than that of the unbound gas ($\simeq 0.52M_{\star}$).
The energy at which the peak mass return rate (Figure~\ref{fig:massreturnrate}) occurs is $E\simeq -1.64\Delta\epsilon$, or $\Xi = 1.64$.  If we assume the debris follows ballistic orbits, the energy distribution presented in Figure~\ref{fig:debris_e} may be translated into a mass fallback rate as a function of time (Figure~\ref{fig:massreturnrate}).  The mass return rate peaks earlier by a factor of $|\Xi|^{-1.5}\simeq 0.5$ than the traditional order of magnitude estimate predicts, 
and the maximal fallback rate is greater by a factor $\simeq 2.5$.

After the bound debris goes out through its orbital apocenter and returns to the region close to the BH, it undergoes multiple shocks (initially near pericenter and apocenter), and finally forms an eccentric flow, as illustrated in Figure~\ref{fig:densitydistribution} (see Section~\ref{subsec:debris} for more details)\footnote{The density fluctuations visible in newly-returning matter are likely numerical artifacts; they are erased by the first shock the gas encounters and have no subsequent influence.} Near pericenter, strengthening vertical gravity and orbital convergence compress the returning debris, creating a ``nozzle" shock (as predicted by \citealt{EvansKochanek1989}) visible at $t\gtrsim 0.5 t_{0}$.  Over time, the shocked gas becomes hotter and thicker. On the way to apocenter, the gas cools adiabatically\footnote{We ignore here the effect of recombination as this energy is negligible compared with the orbital energy: even at apocenter, the ratio of gas kinetic energy to recombination energy is $\sim (GM_*/R_*)/E_{\rm recomb} \sim O(100)$.}.  Near apocenter, the previously-shocked outgoing debris collides with fresh incoming debris, creating another shock (the ``apocenter" shock).  Previously-shocked gas close to the orbital plane is deflected inward toward the BH, while the portion farther from the plane is deflected above and below the incoming stream (see Section~\ref{subsection:poloidal}). These deflections broaden the angular momentum distribution. A small part of the debris loses enough angular momentum that it acquires a pericenter smaller than the star's pericenter.  Other gas gains angular momentum, which results in the nozzle shock-front gradually extending to larger and larger radii.

At $t \simeq (1-2)t_{0}$, the debris in the apocenter region undergoes a dramatic transition in shape, from well-defined incoming and outgoing streams to an extended eccentric accretion flow. By the time the return rate of the newly incoming debris declines, the mass that had arrived earlier becomes large enough to significantly disrupt the newly incoming debris' orbit. The space inside the apocenter region is then quickly filled with gas.
The outcome is an extended eccentric accretion flow ($e\simeq 0.4-0.5$ at $t\simeq 3t_{0}$), most of whose mass resides at radii $\sim 10^3 - 10^4 r_{\rm g}$ (see Figure~\ref{fig:massenergybudget}). At $t\simeq 3t_{0}$, only $2-3\times 10^{-2}\Msol$
can be found inside $2r_{\rm p}$.  By contrast, 
from $t \approx t_0$ onward, nearly all the thermal energy is found at small radii, a condition that has consequences for the time-dependence of escaping radiation. Throughout the volume occupied by the debris, radiation pressure dominates: it is generally larger than gas pressure by a factor $\sim 10^3 - 10^4$.

\begin{figure*}
\includegraphics[width=4.4cm]{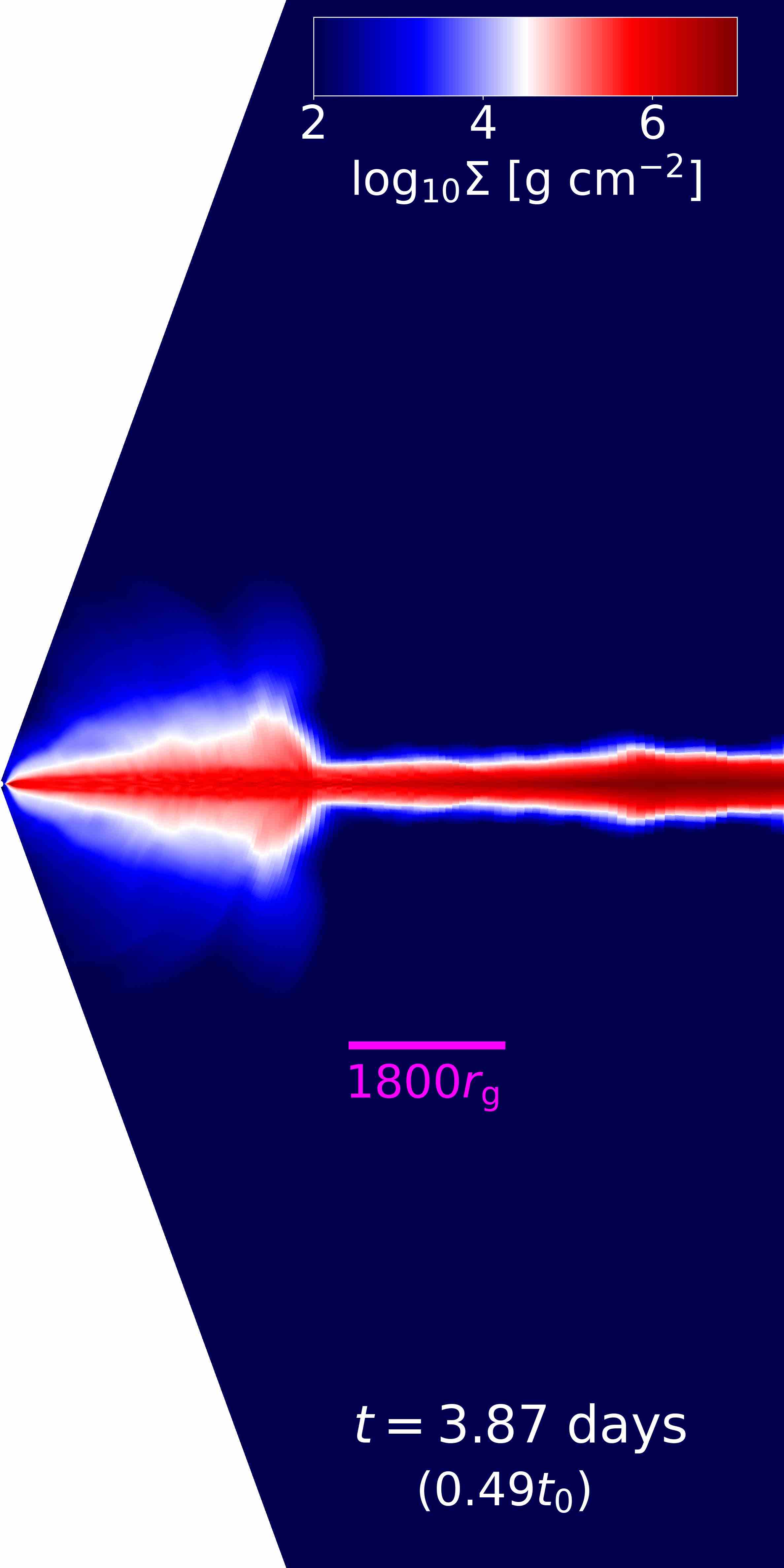}
\includegraphics[width=4.4cm]{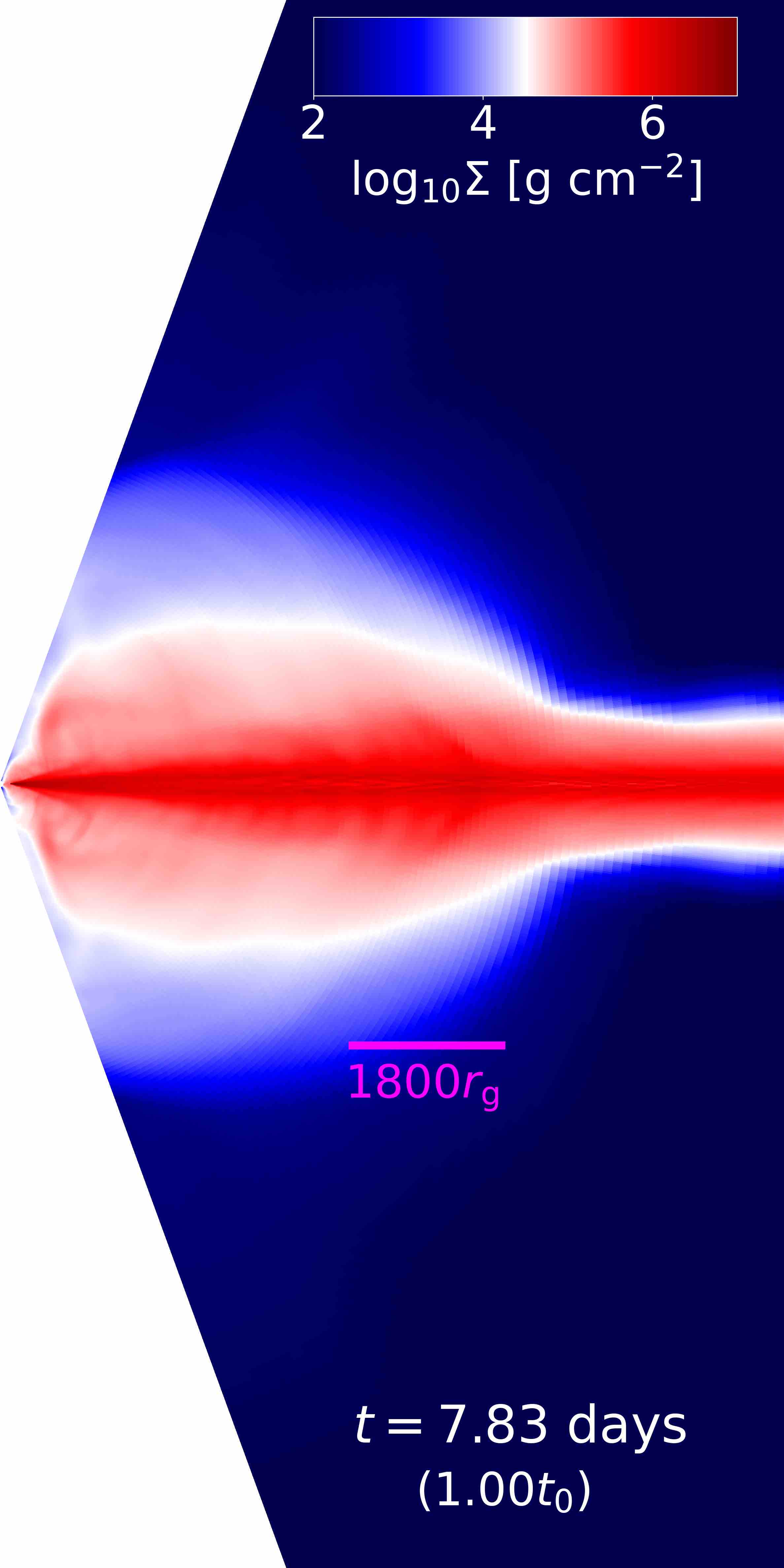}
\includegraphics[width=4.4cm]{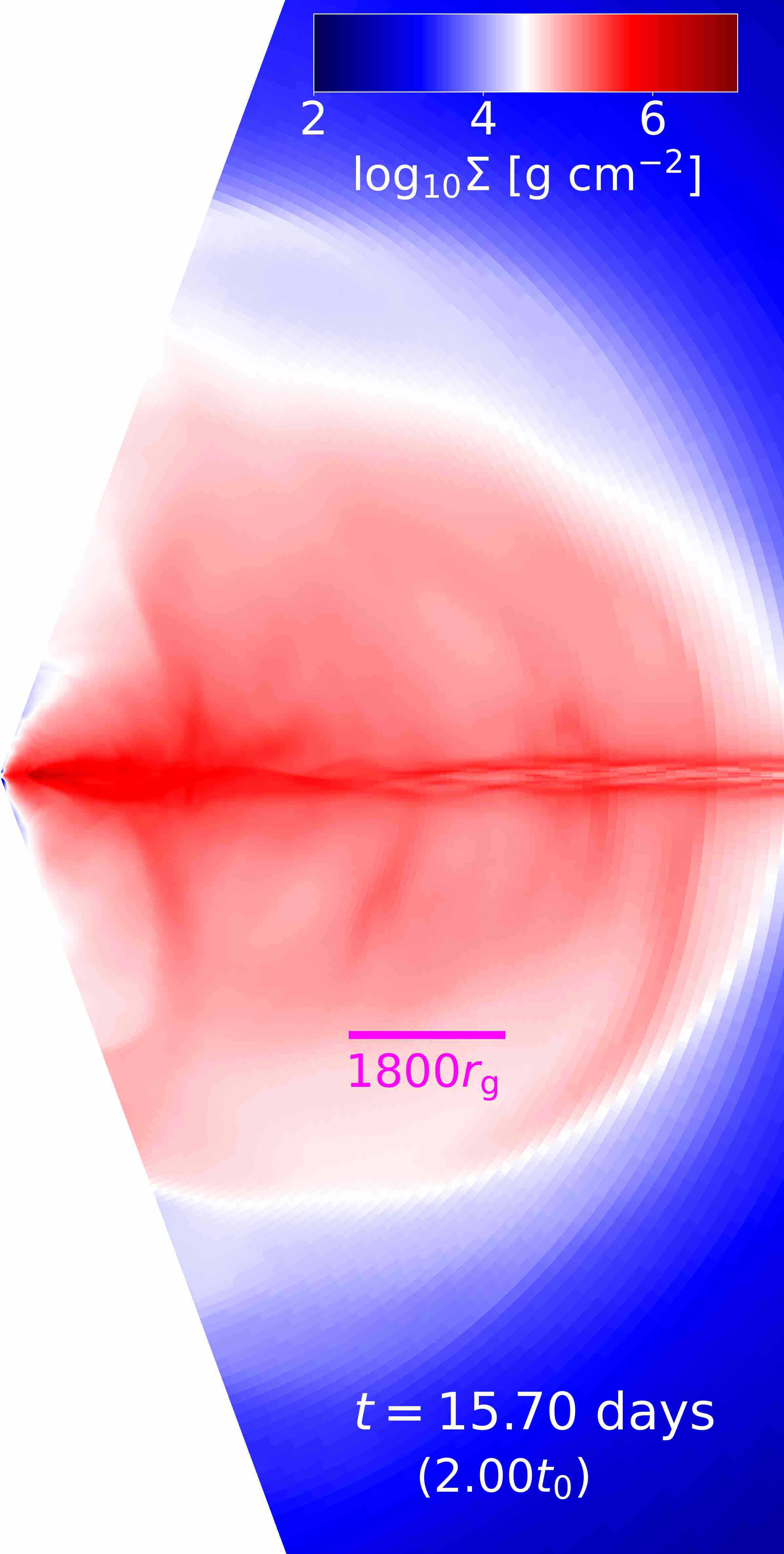}
\includegraphics[width=4.4cm]{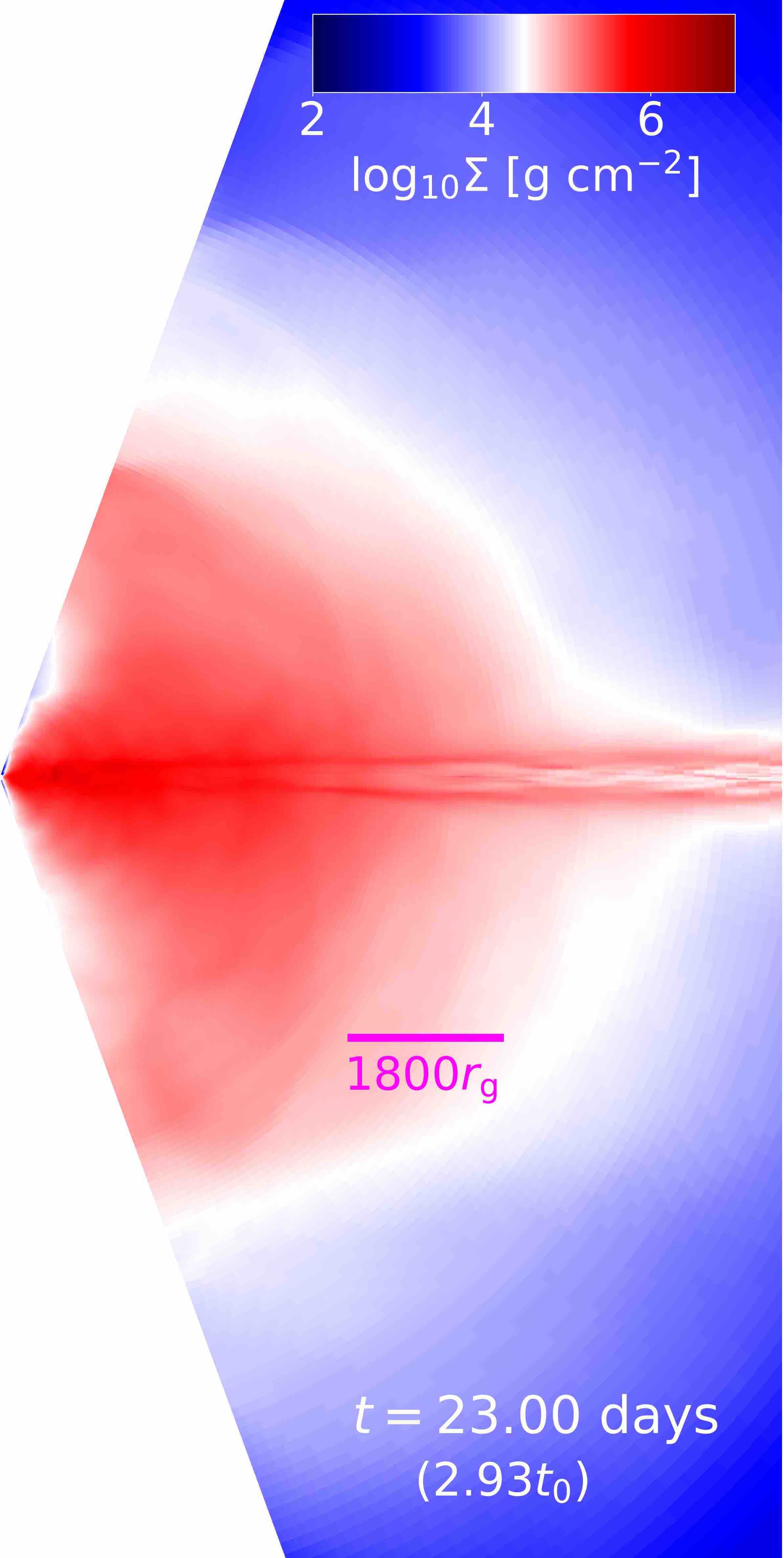}
	\caption{The azimuthally integrated density distribution at $t/t_{0}=0.5$, 1, 2, and 3.  }
	\label{fig:verticaldensity}
\end{figure*}

In other words, \textit{circularization is not prompt}: the flow retains significant eccentricity, and the great majority of the gas remains at a distance $10^3 - 10^4 r_{\rm g} \gg r_{\rm p}$ even after several characteristic timescales.
That this is so can also be seen from another point of view.  At $t\simeq 3t_{0}$,  the total amount of dissipated energy is only 10\% of the  energy, $E_{\rm circ} \equiv G\Mbh/4r_{\rm p}$,  required for the debris to fully ``circularize" into a compact disk on the commonly-expected radial scale of $2r_{\rm p}$. 
Extrapolating this slow energy dissipation rate to late times suggests that true ``circularization" would take a few tens of $t_{0}$ (see Figure~\ref{fig:circularization}).
As we will discuss in detail in Section~\ref{dis:circularization}, the total dissipation rate is roughly constant with time from $t \simeq 0.5t_0$ until the end of our simulation at $t\simeq 3t_0$.  Thus, there is no runaway increase in dissipation of the sort suggested by \citet{SteinbergStone2022}.

\begin{figure*}
\includegraphics[width=5.7cm]{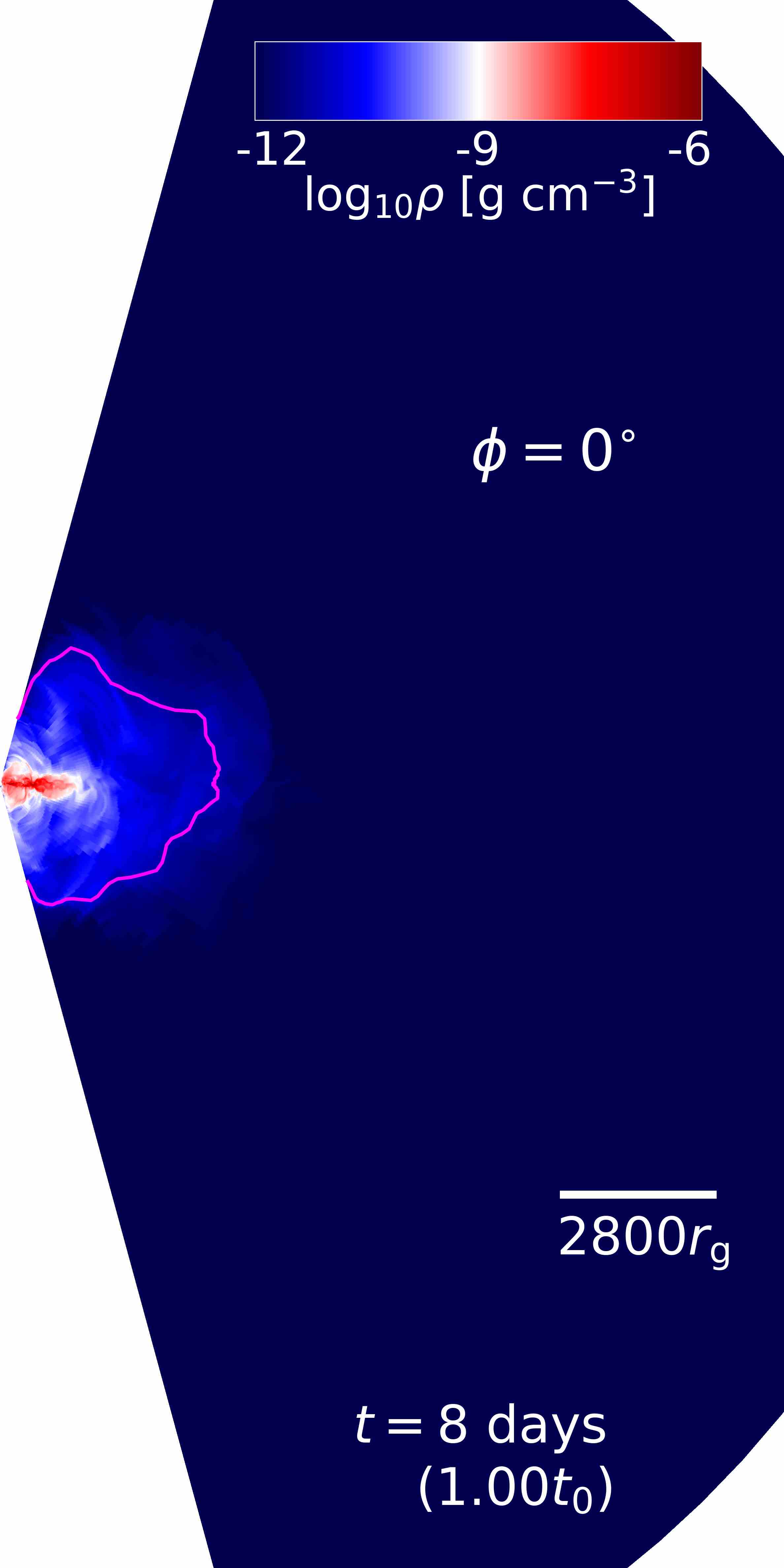}
\includegraphics[width=5.7cm]{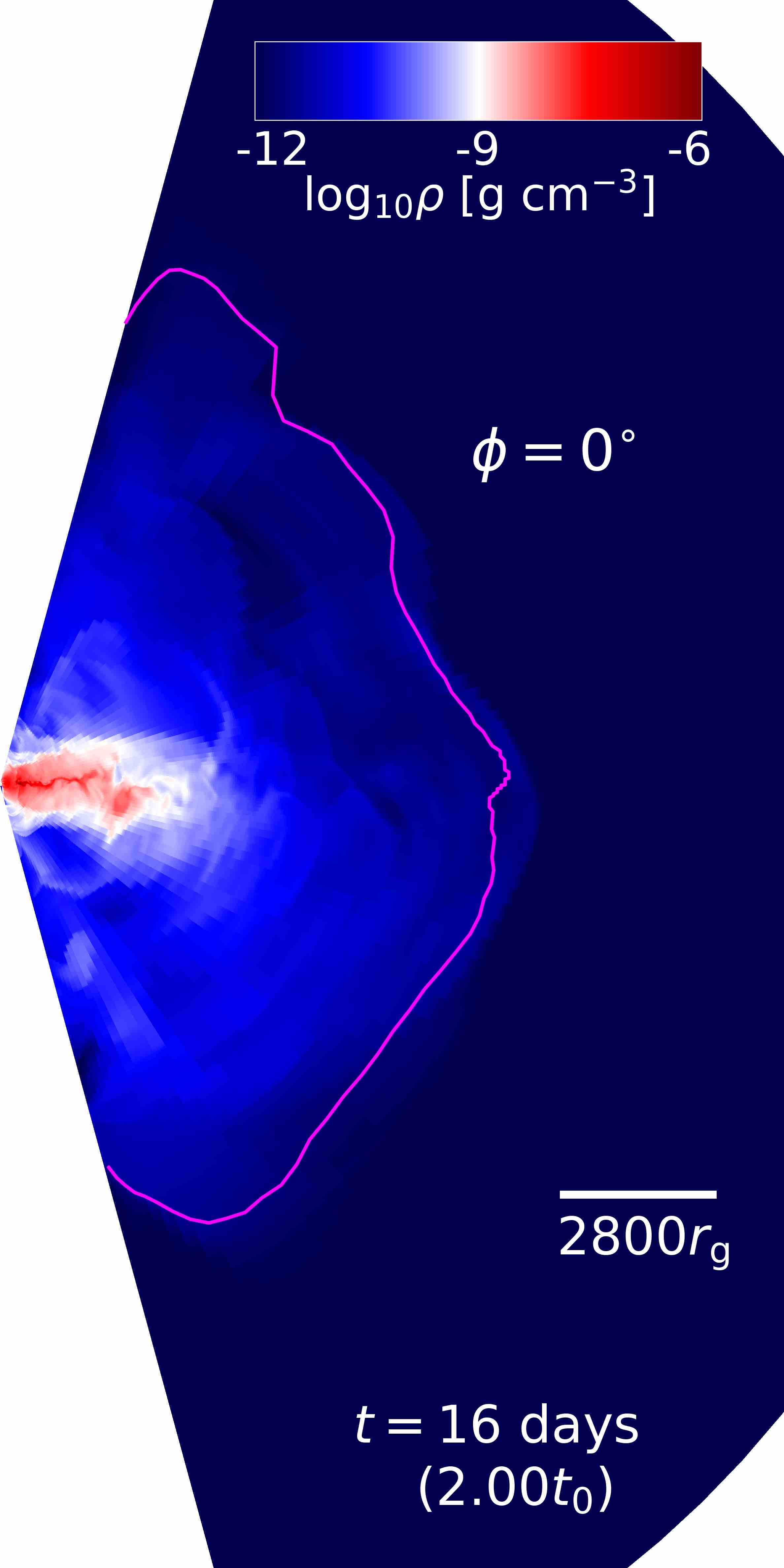}
\includegraphics[width=5.7cm]{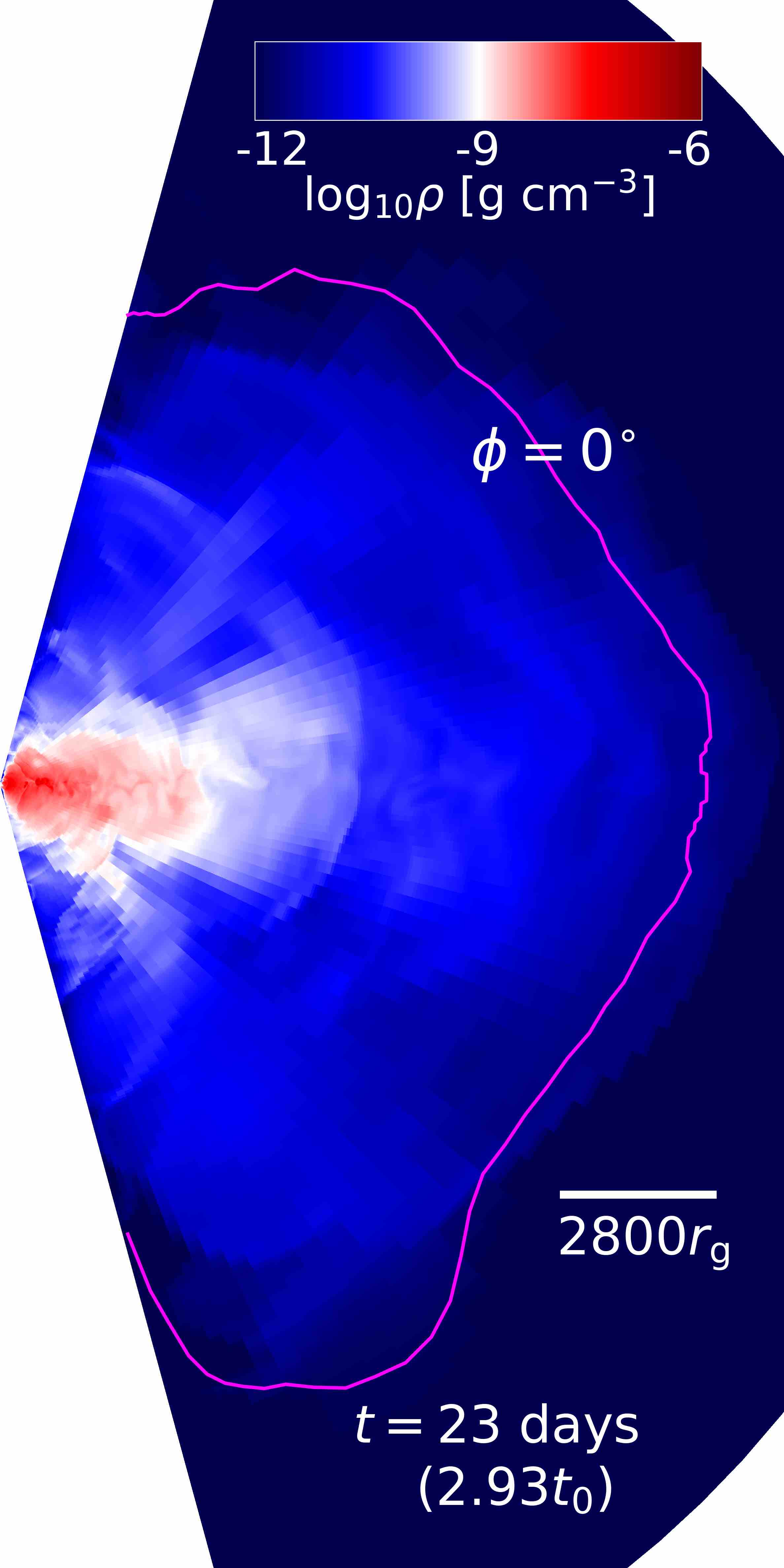}\\
\includegraphics[width=5.7cm]{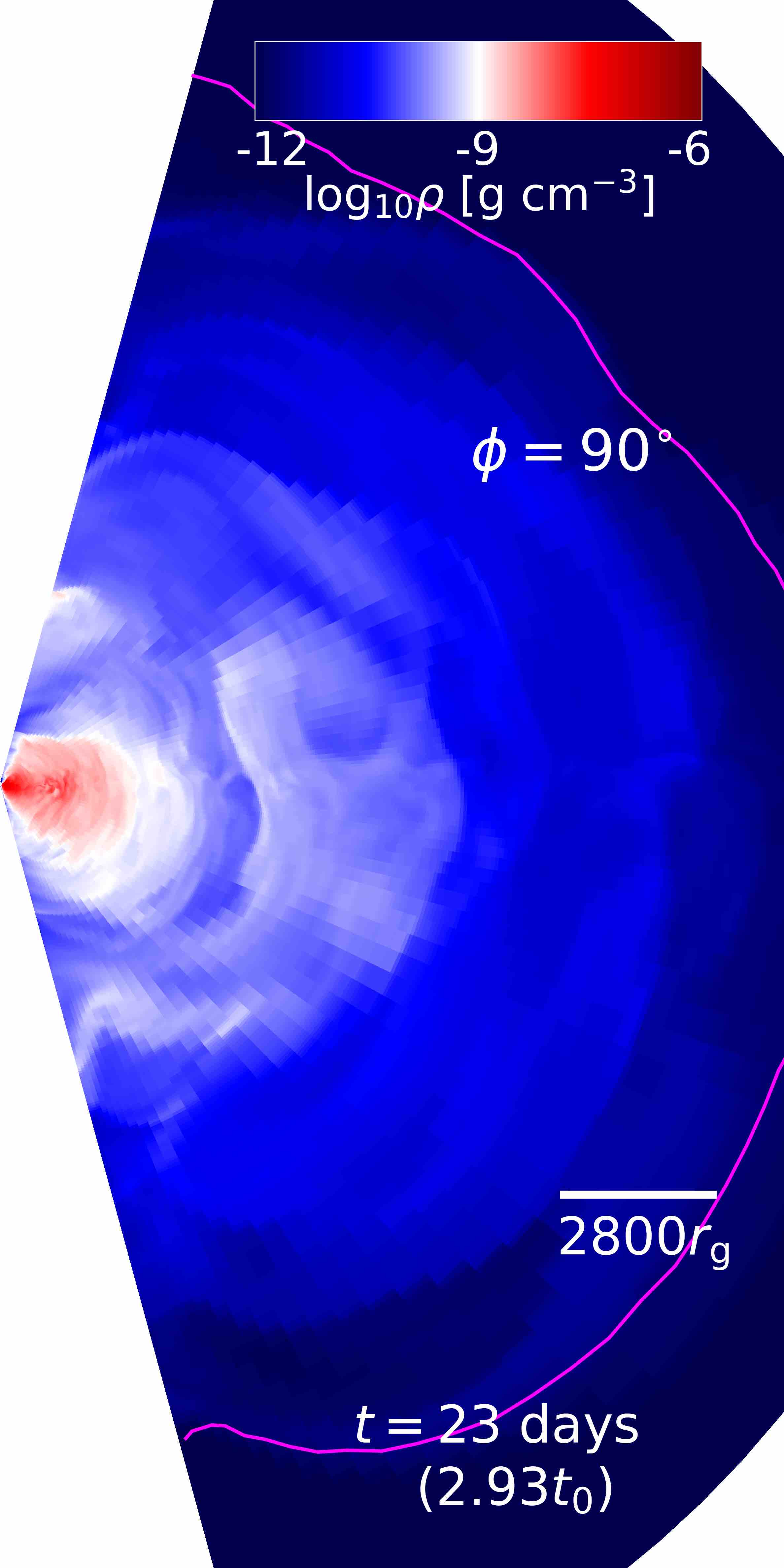}
\includegraphics[width=5.7cm]{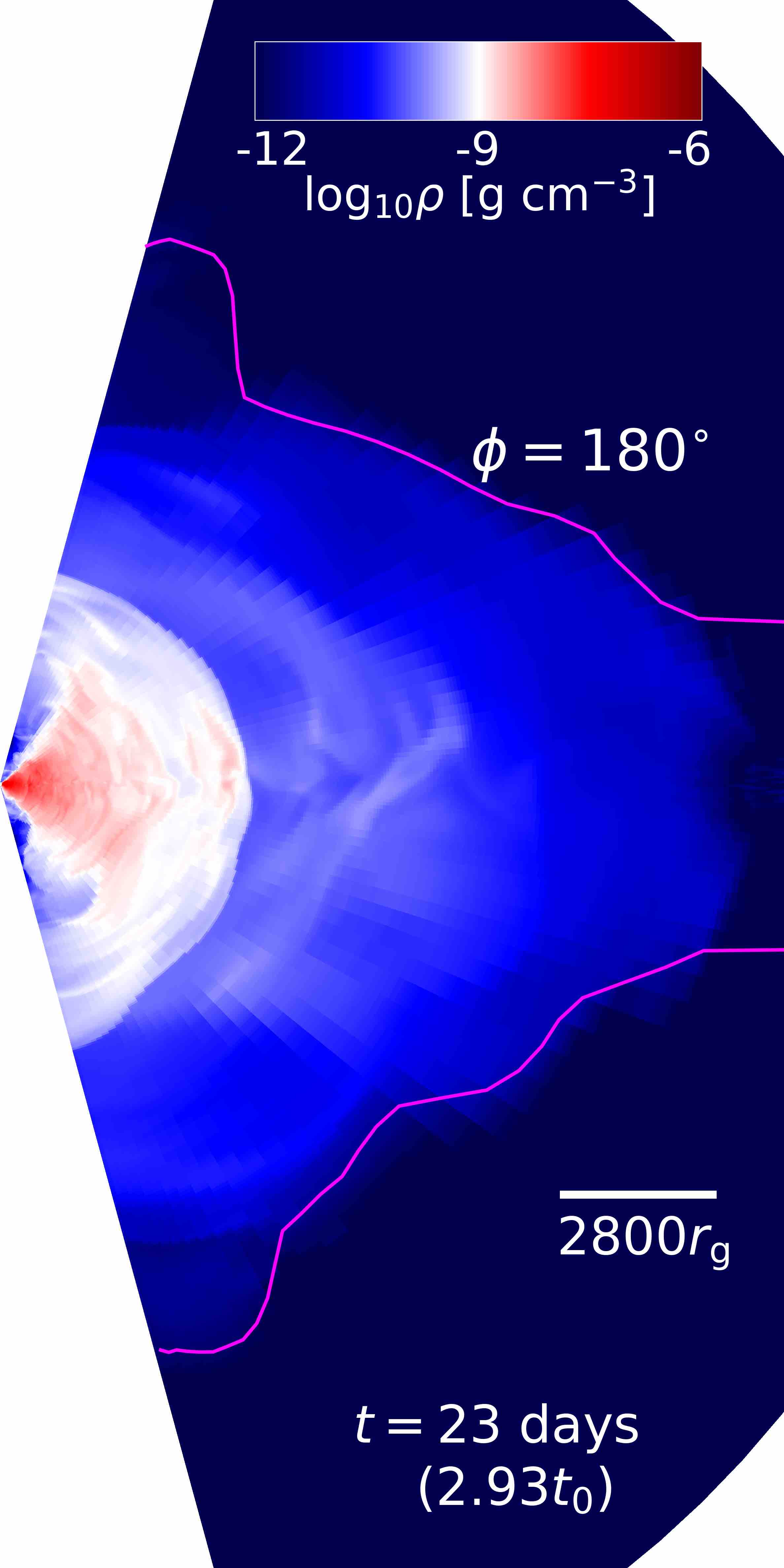}
\includegraphics[width=5.7cm]{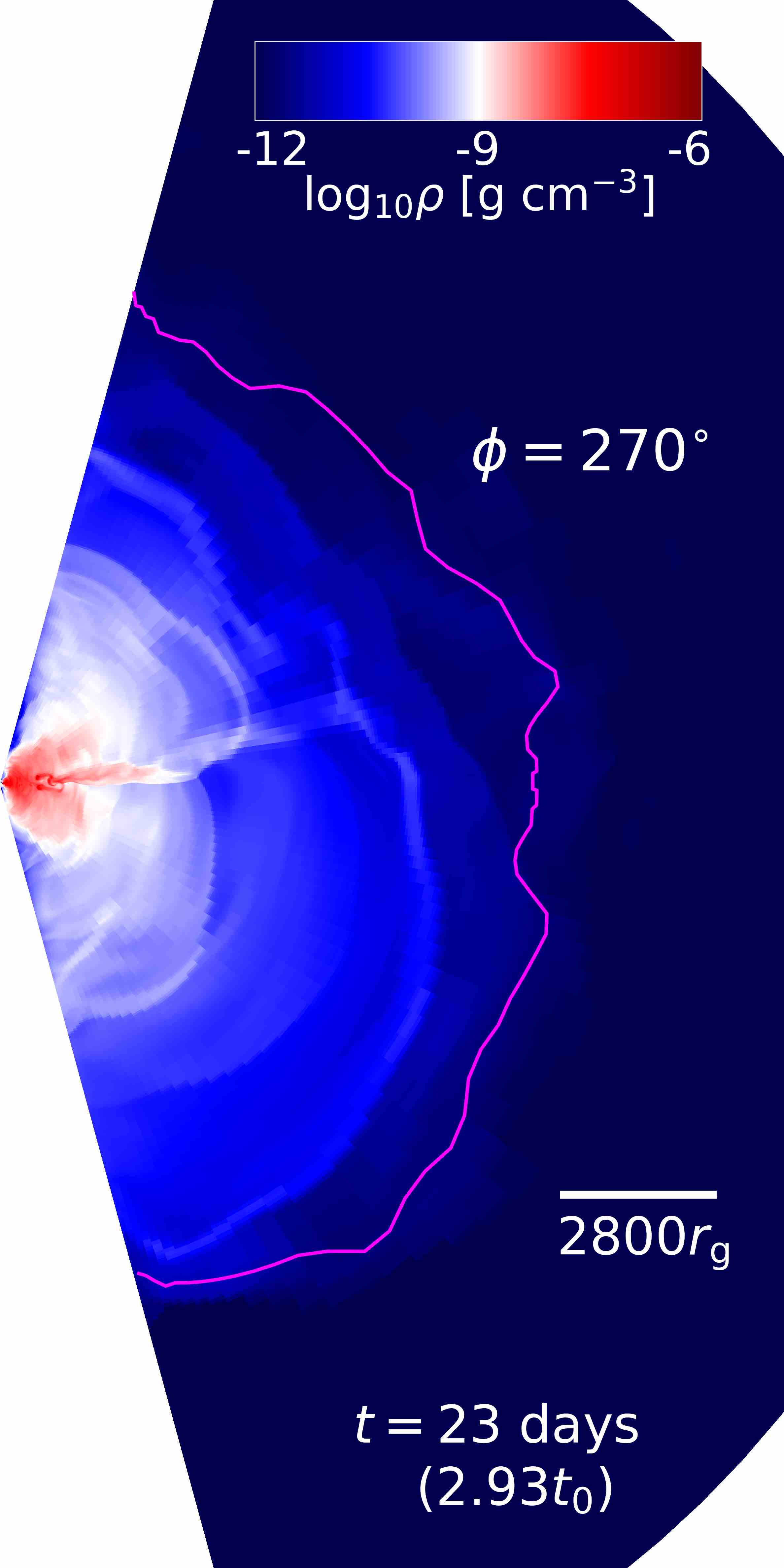}
	\caption{The location of the thermalization photosphere (magenta curves) as seen by a distant observer plotted over the density distribution at $\phi=0$ at $t/t_{0}=1$, $2$, $3$ (\textit{top} panels) and at $\phi=90^{\circ}$, $180^{\circ}$ and $270^{\circ}$ at $t/t_{0}=3$ (\textit{bottom panels}).  We define the thermalization optical depth as $\sqrt{\tau_{\rm t} \tau_{\rm ff}}$, for $\tau_{\rm T}$ ($\tau_{\rm ff}$) the Thomson (absorption) optical depths integrated radially inwards from the outer boundary.}
	\label{fig:photosphere_t}
\end{figure*}

The shocked gas expands outward quasi-symmetrically (Figure~\ref{fig:verticaldensity}).  Because the intrinsic binding energy of the debris is much smaller than $E_{\rm circ}$, the dissipated energy is large enough to be comparable to the specific orbital energy.  As a result, the expanding material is marginally bound. The radial expansion speed of the gas near the photosphere ($r\simeq 7000-10000r_{\rm g}$, Figure~\ref{fig:photosphere_t}) at $t\simeq 3t_{0}$ is $0.005-0.01c\simeq 1500 - 3000$ km/s; the associated specific energy is $\lesssim 10^{-4}c^2$, comparable to the intrinsic energy scale, $\Delta E$.

Although we do not incorporate radiation transfer into the simulation, we estimate the luminosity in post-processing (see \S~\ref{sub:radiation}).  The bolometric luminosity rises to $(6-7)\times 10^{43}$erg/s in $t\simeq t_{0}$; its effective temperature on the photosphere is generally close to $\sim 2\times 10^5$~K.

\begin{figure*}\centering
\includegraphics[width=5.8cm]{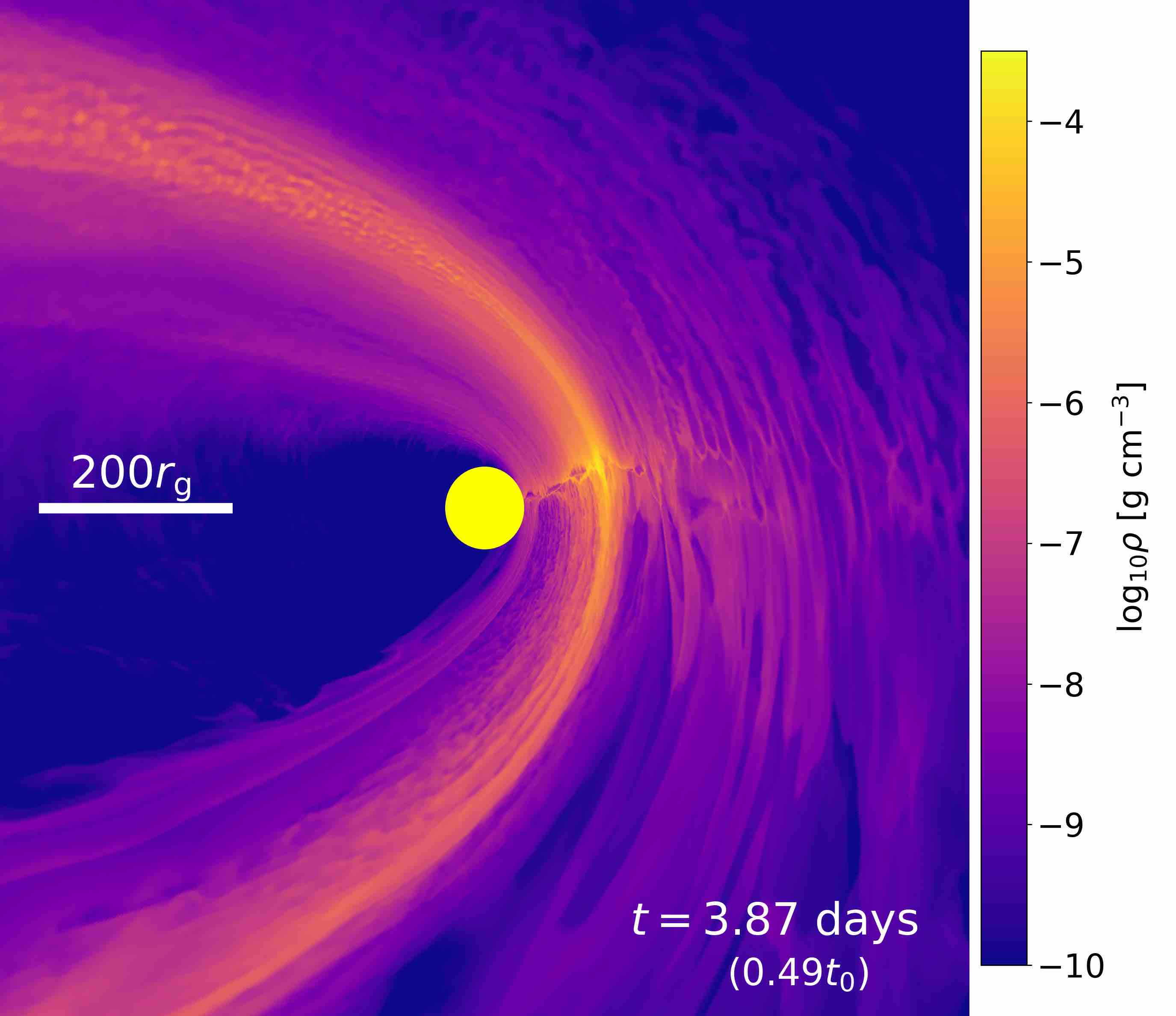}
\includegraphics[width=5.8cm]{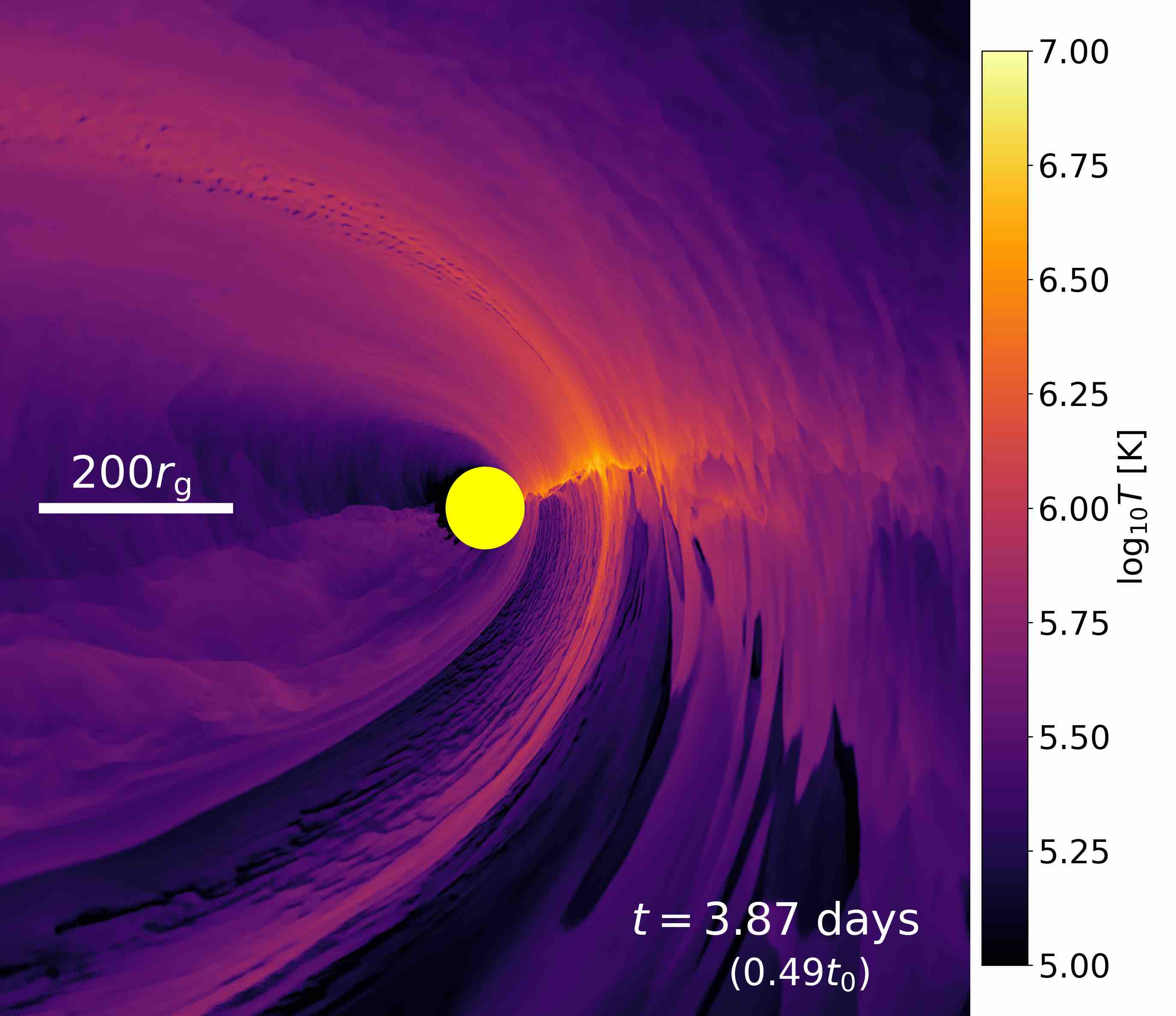}\\
\includegraphics[width=5.8cm]{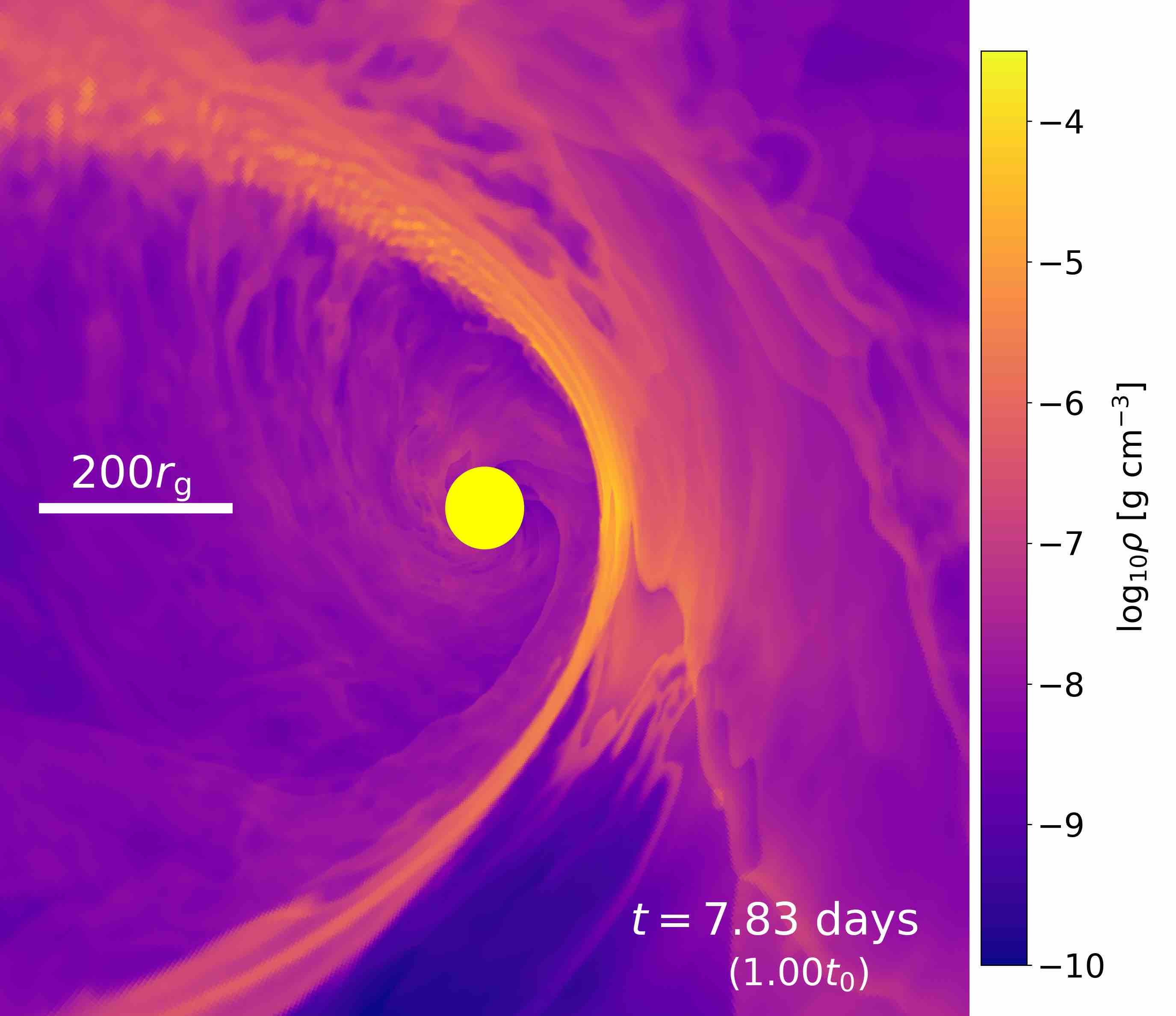}
\includegraphics[width=5.8cm]{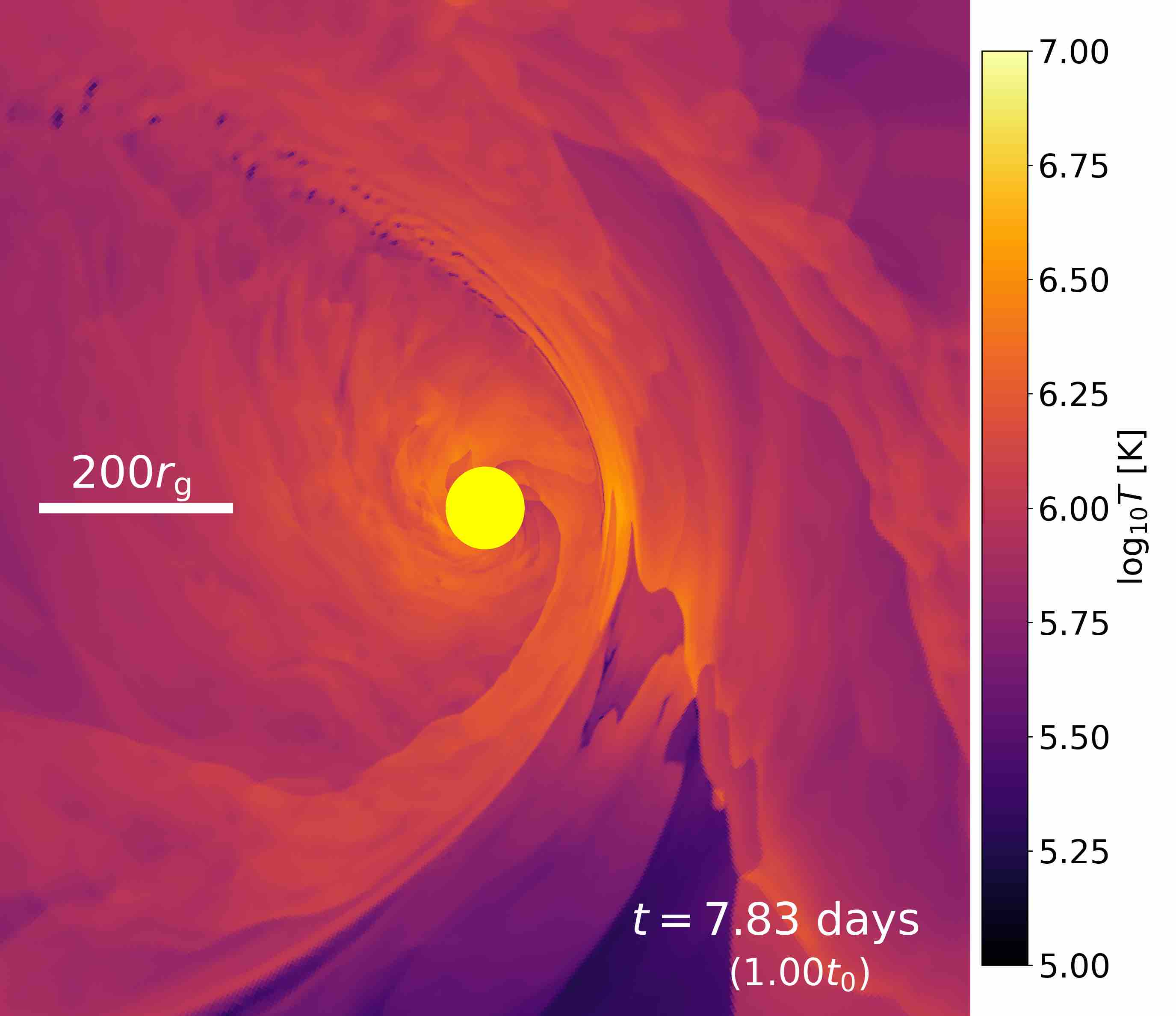}\\
\hspace{0.05in}\includegraphics[width=5.8cm]{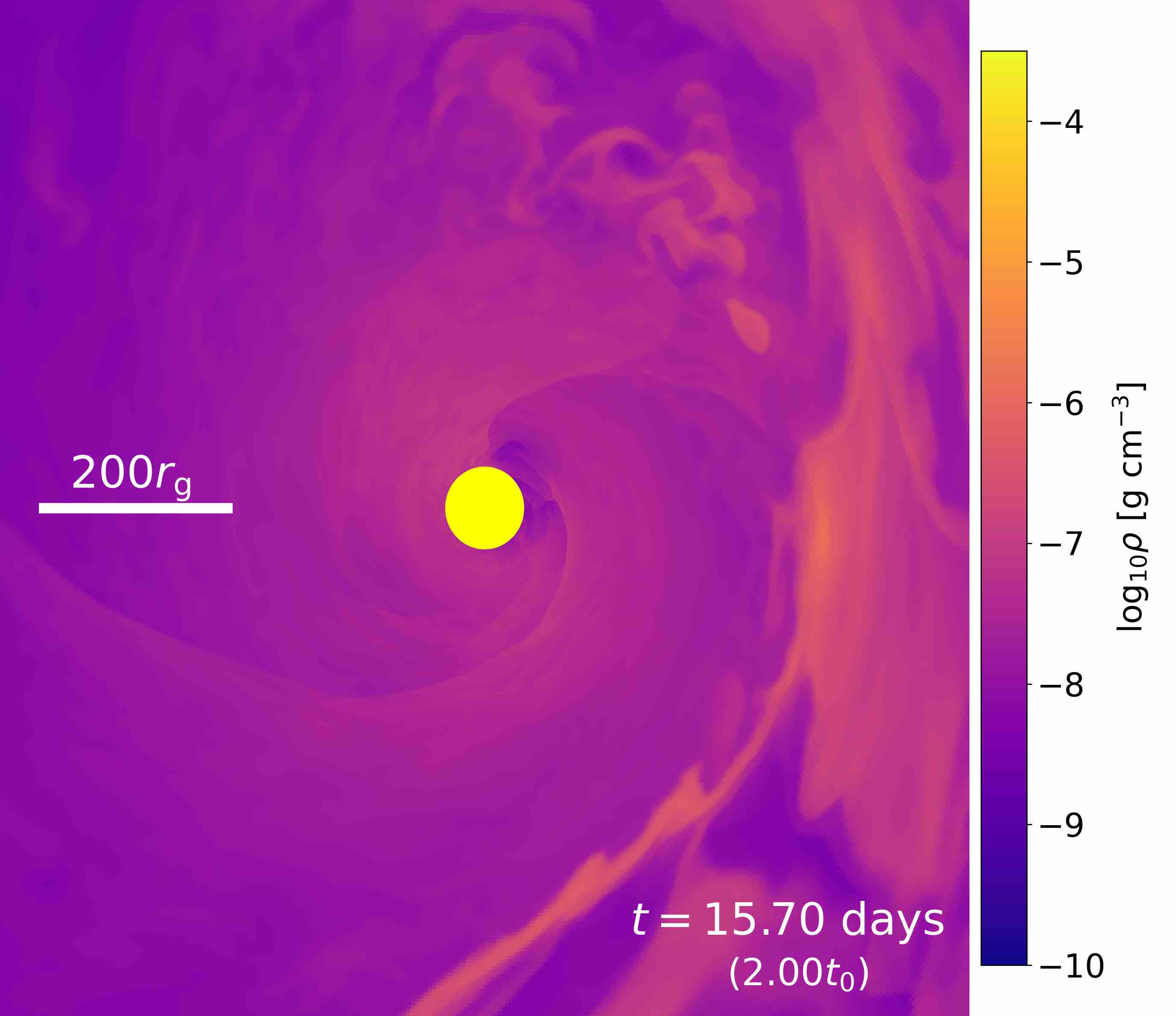}\hspace{-0.03in}
\includegraphics[width=5.8cm]{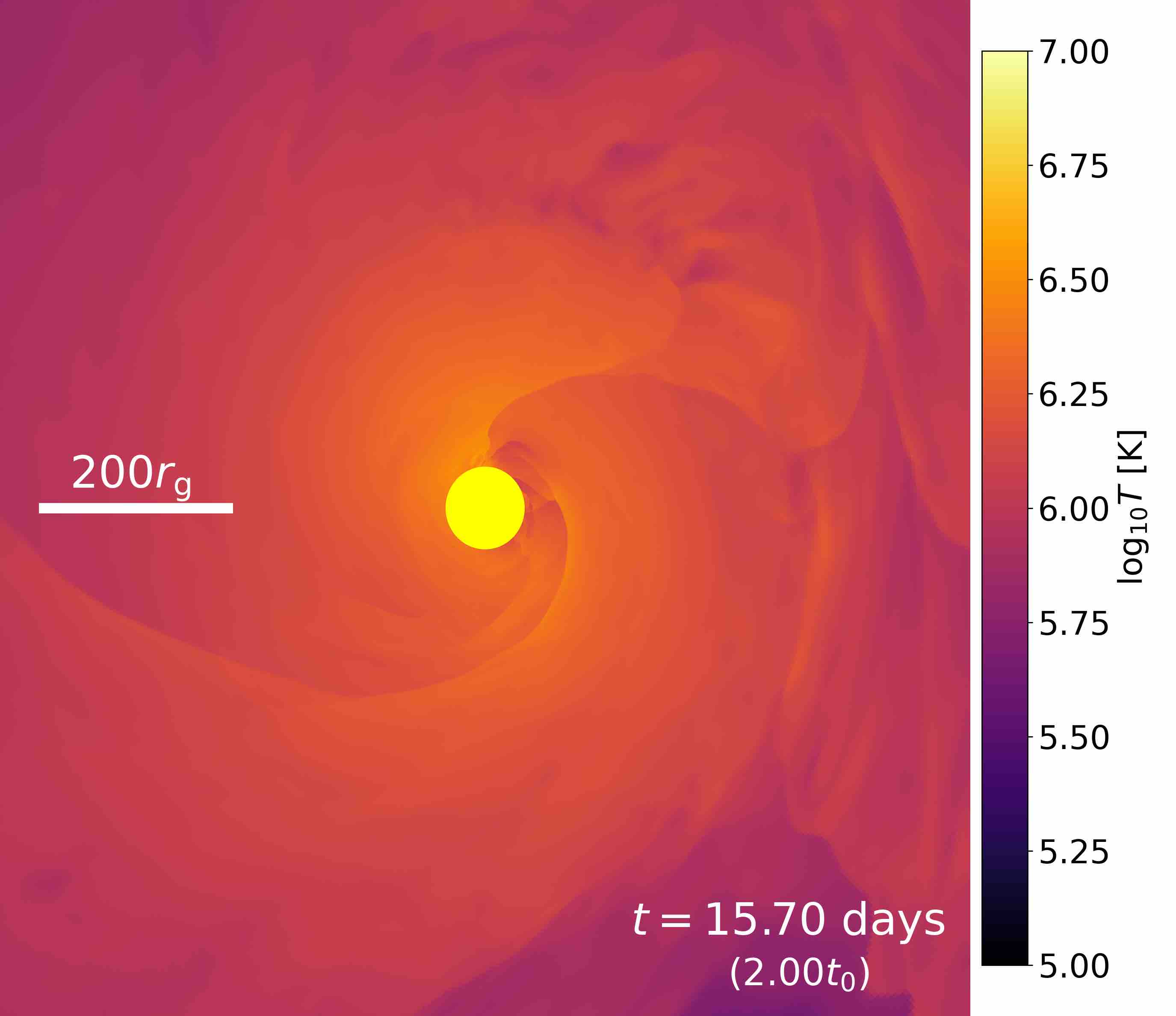}\\
\hspace{0.05in}\includegraphics[width=5.8cm]{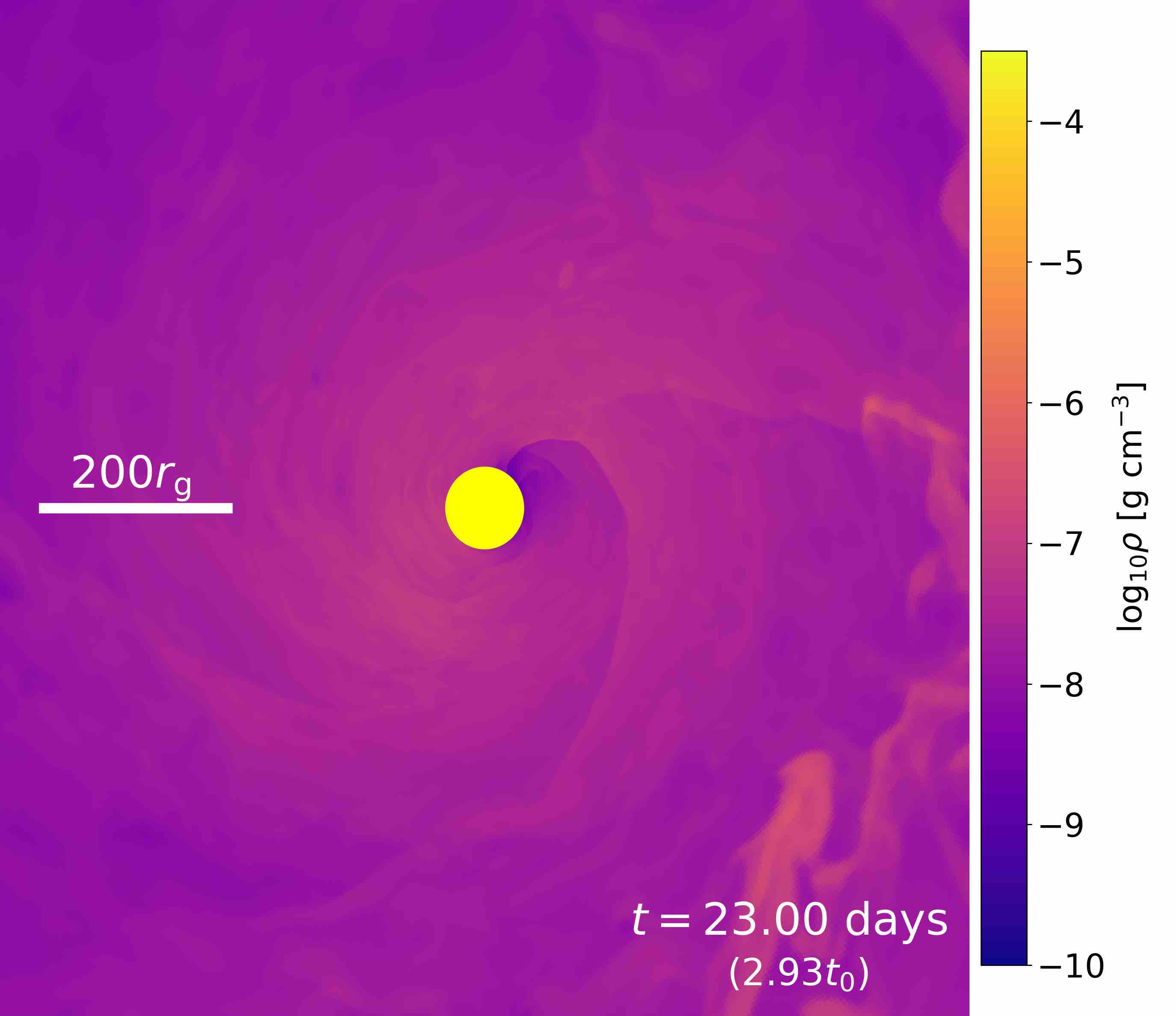}\hspace{-0.03in}
\includegraphics[width=5.8cm]{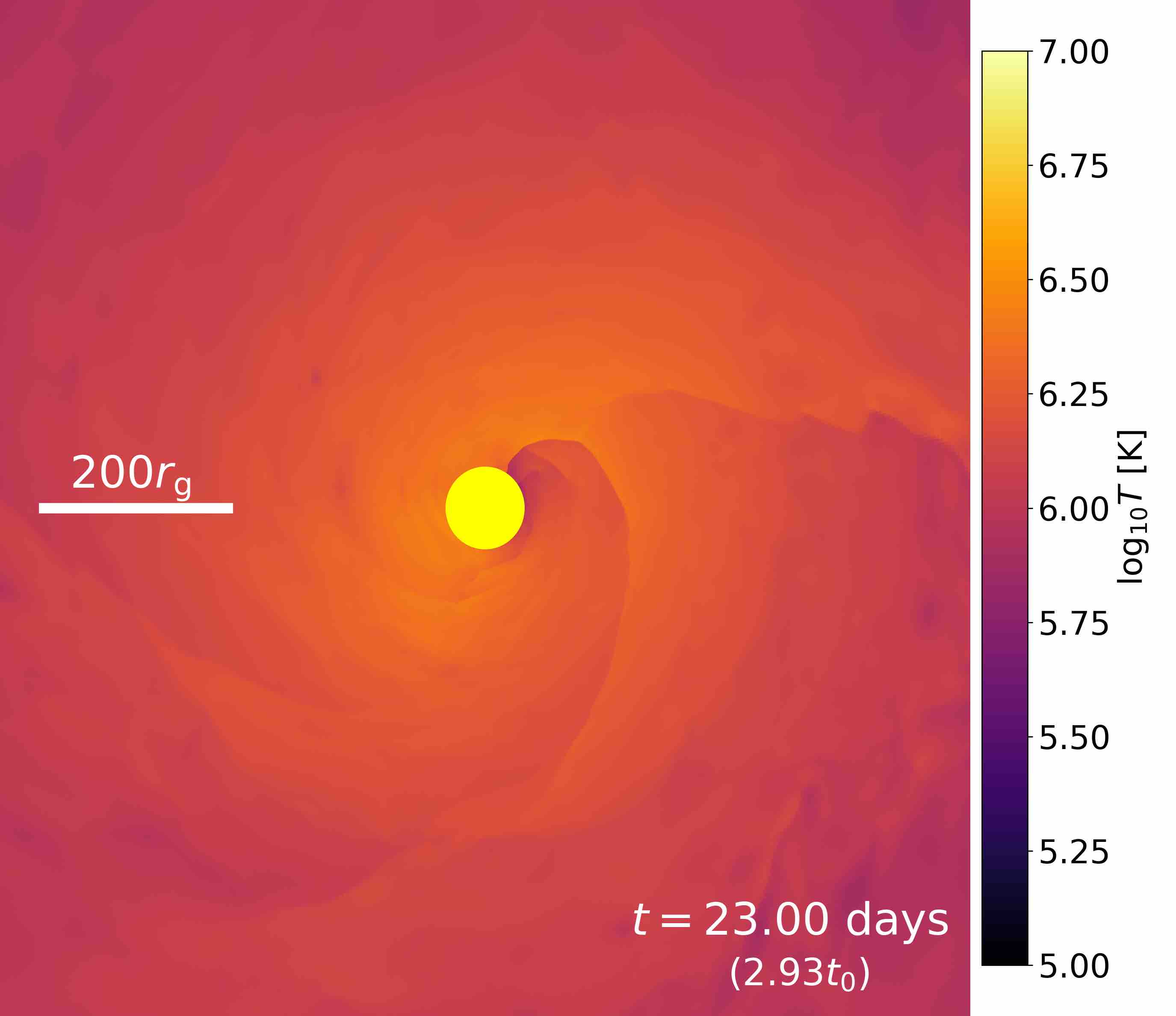}
	\caption{The density (\textit{left} panels) and temperature (\textit{right} panels) distribution  near the black hole (yellow dot) at four different times. Note the stationary spiral shock structure that appears around $2t_0$. }
	\label{fig:nozzleshock}
\end{figure*}

\begin{figure*}
\includegraphics[width=8.6cm,height = 8cm]{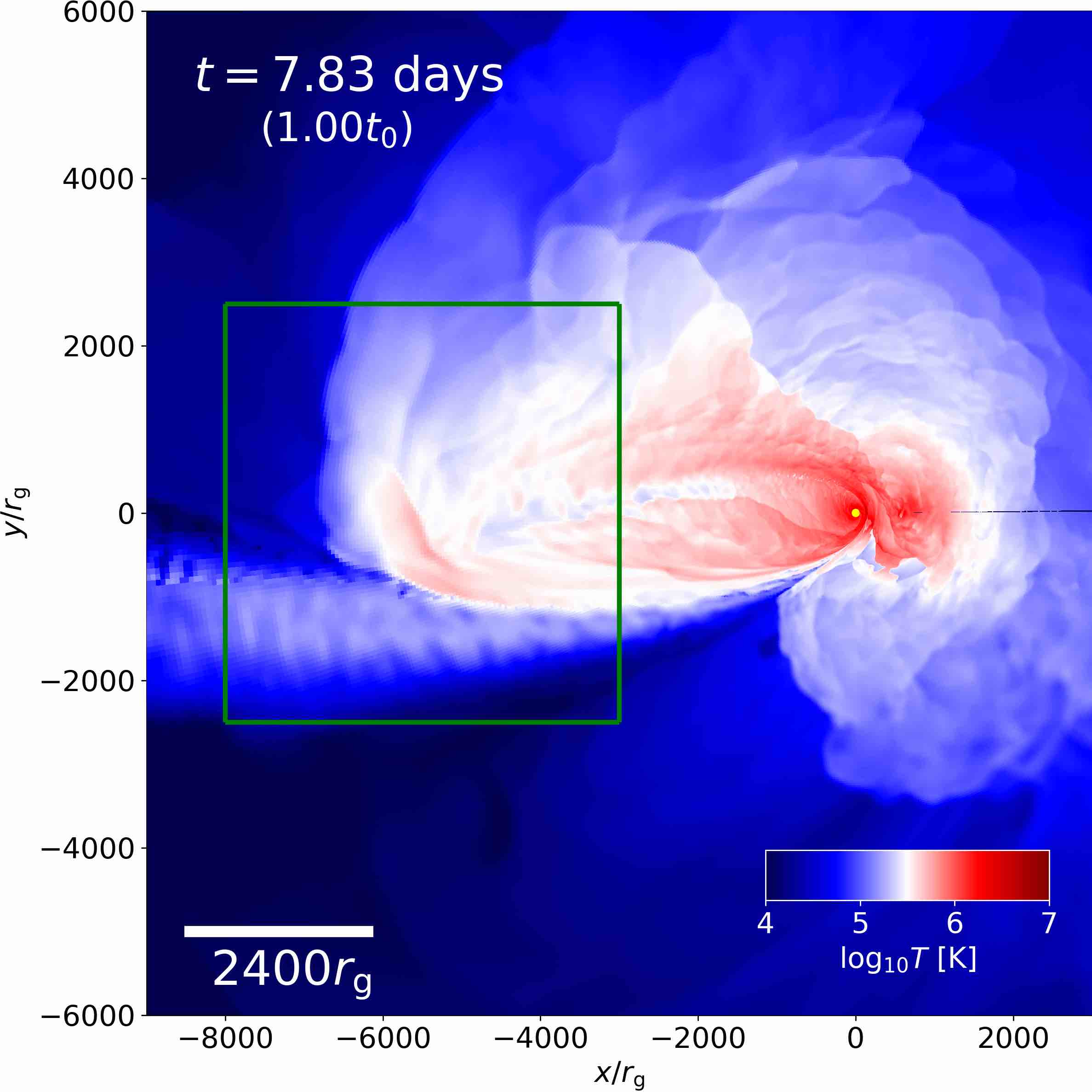}
\includegraphics[width=8.6cm,height = 8cm]{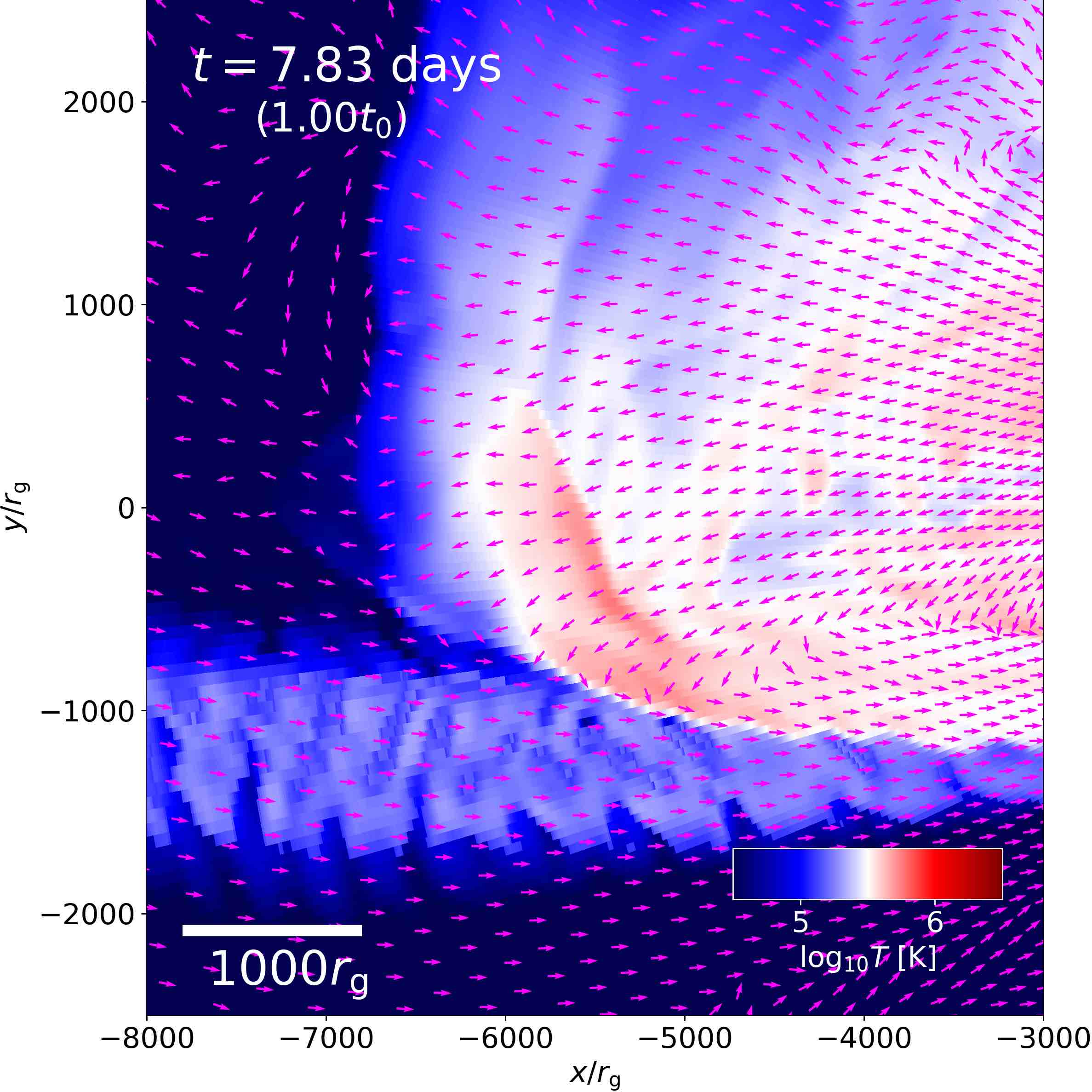}
	\caption{The temperature distribution (\textit{left}) at $t=t_{0}$ in the equatorial plane. The \textit{right} panel shows a zoom-in of the region demarcated by the green box in the \textit{left} panel near where two streams collide. The arrows indicate the motion of gas. Notice the different color scales in the \textit{left} and \textit{right} panels. }
	\label{fig:apocentershock}
\end{figure*}

\begin{figure*}\centering
\includegraphics[width=8.6cm]{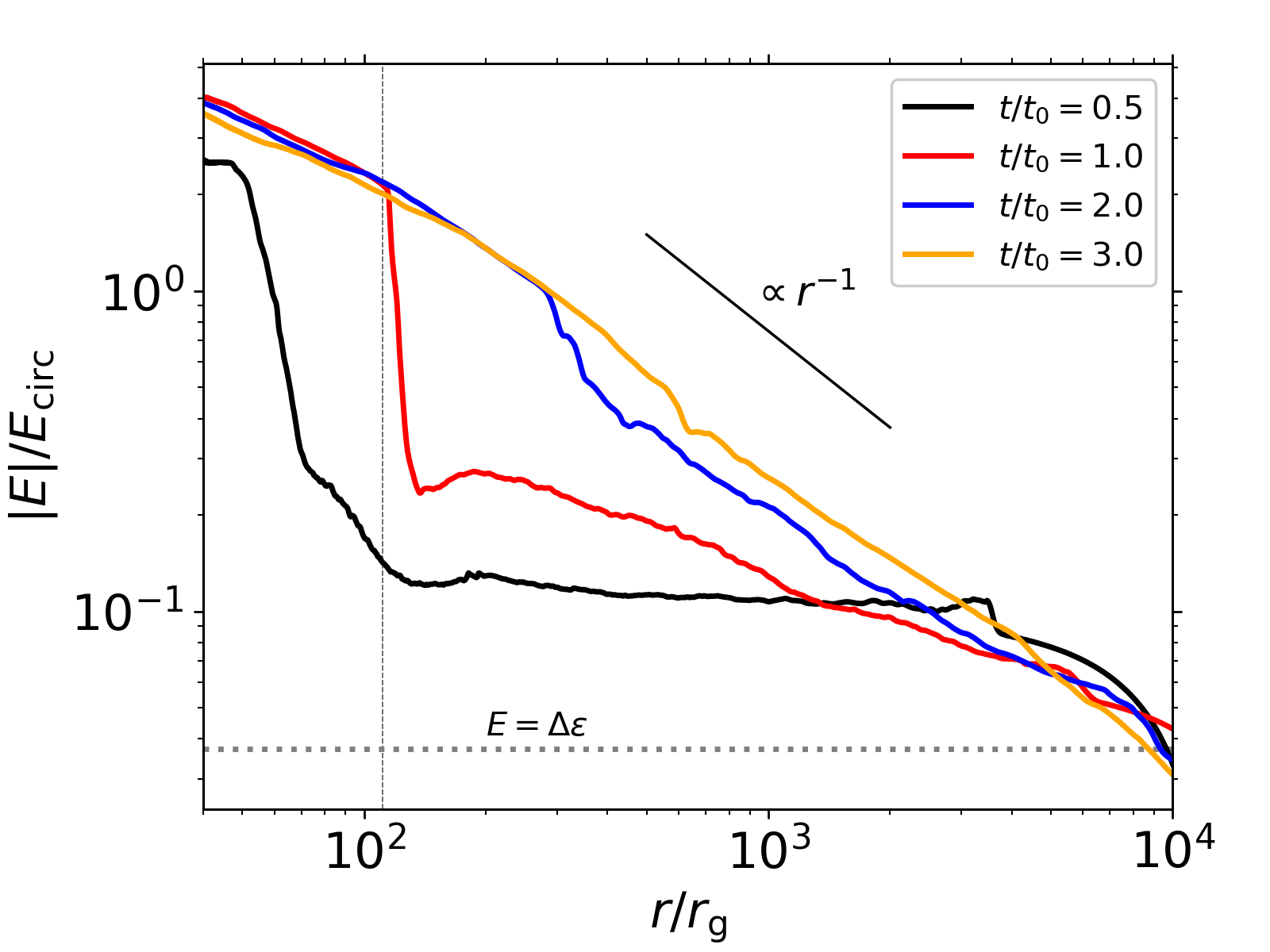}
\includegraphics[width=8.6cm]{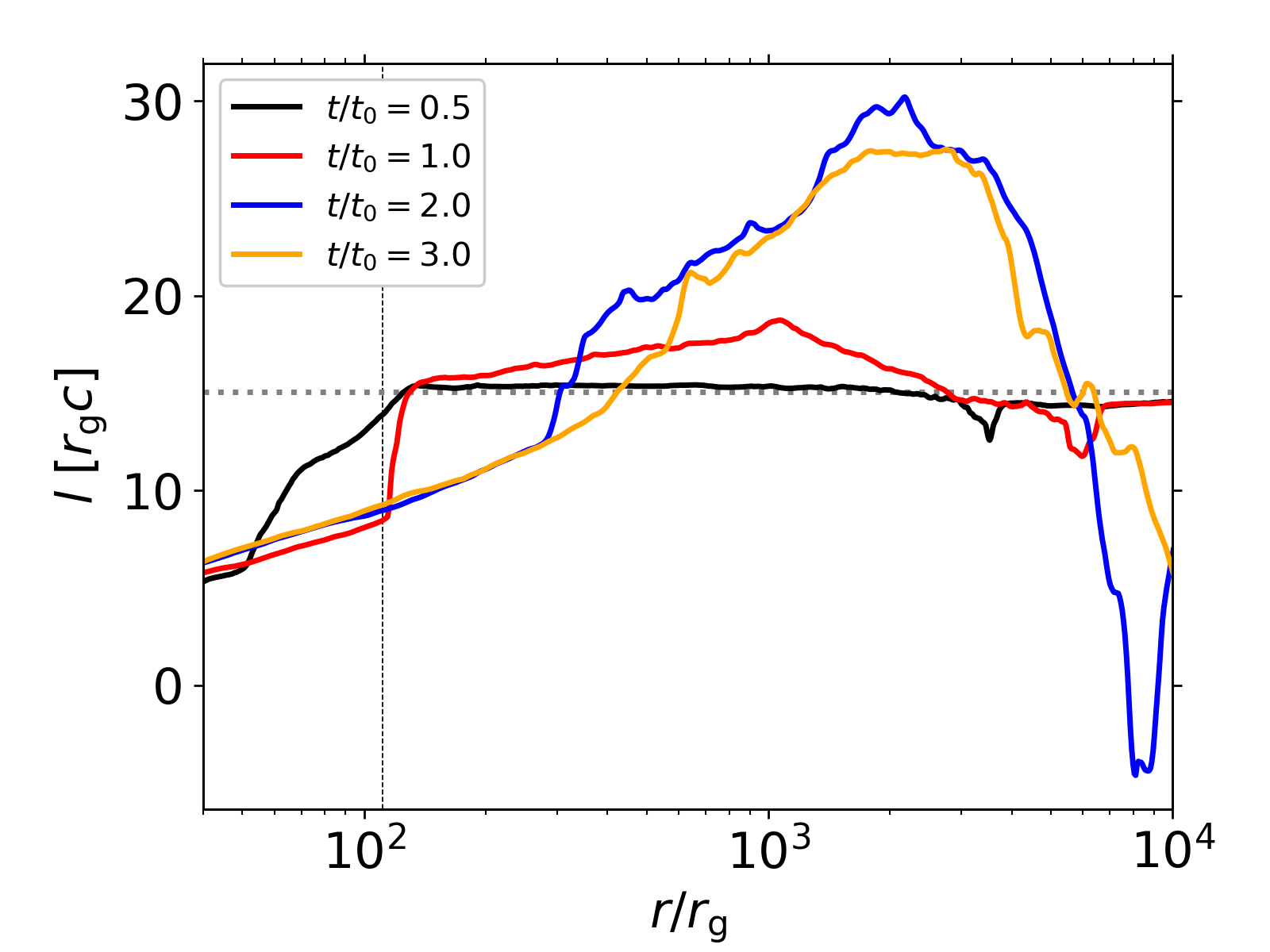}\\
\includegraphics[width=8.6cm]{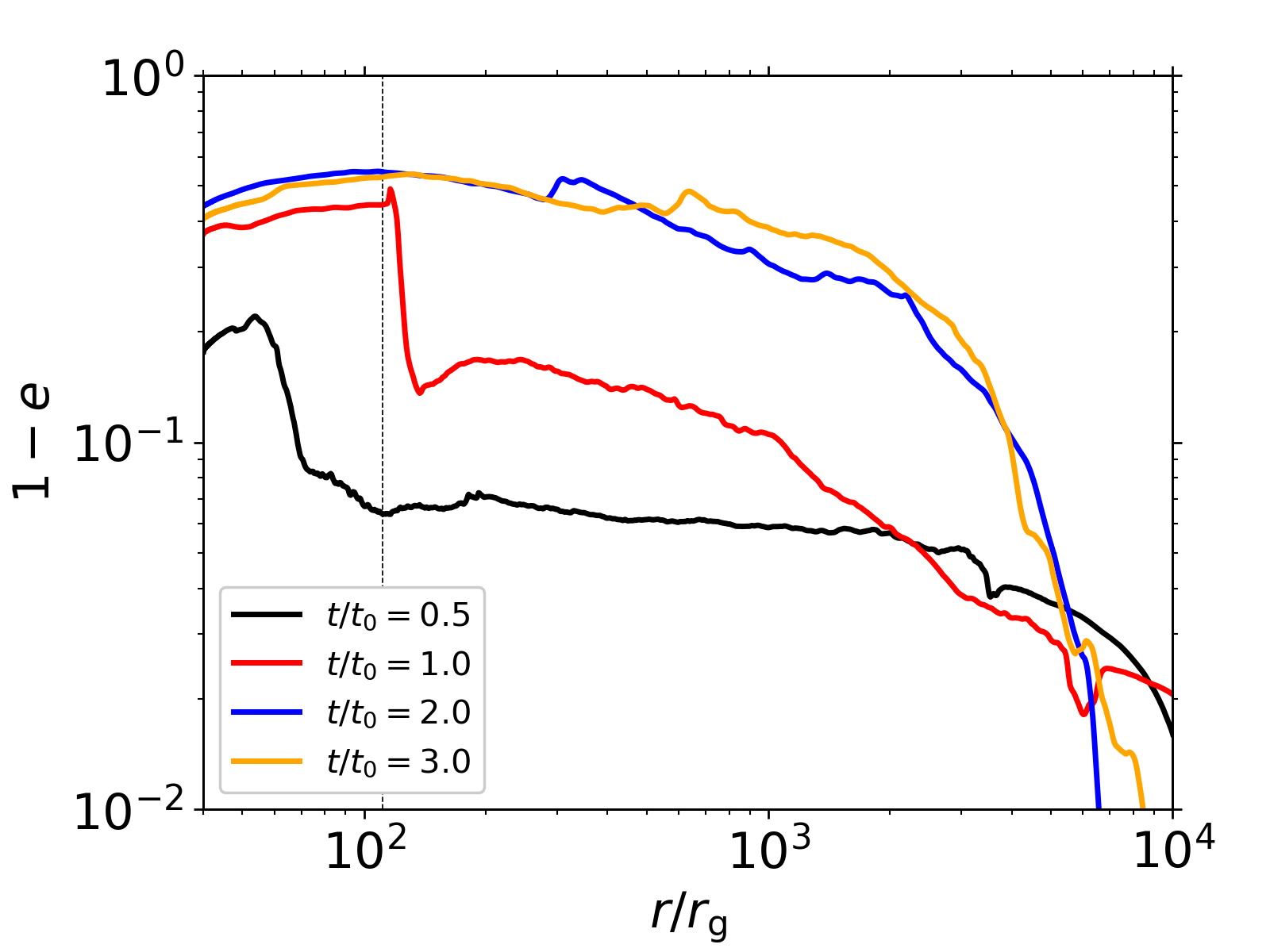}
\includegraphics[width=8.6cm]{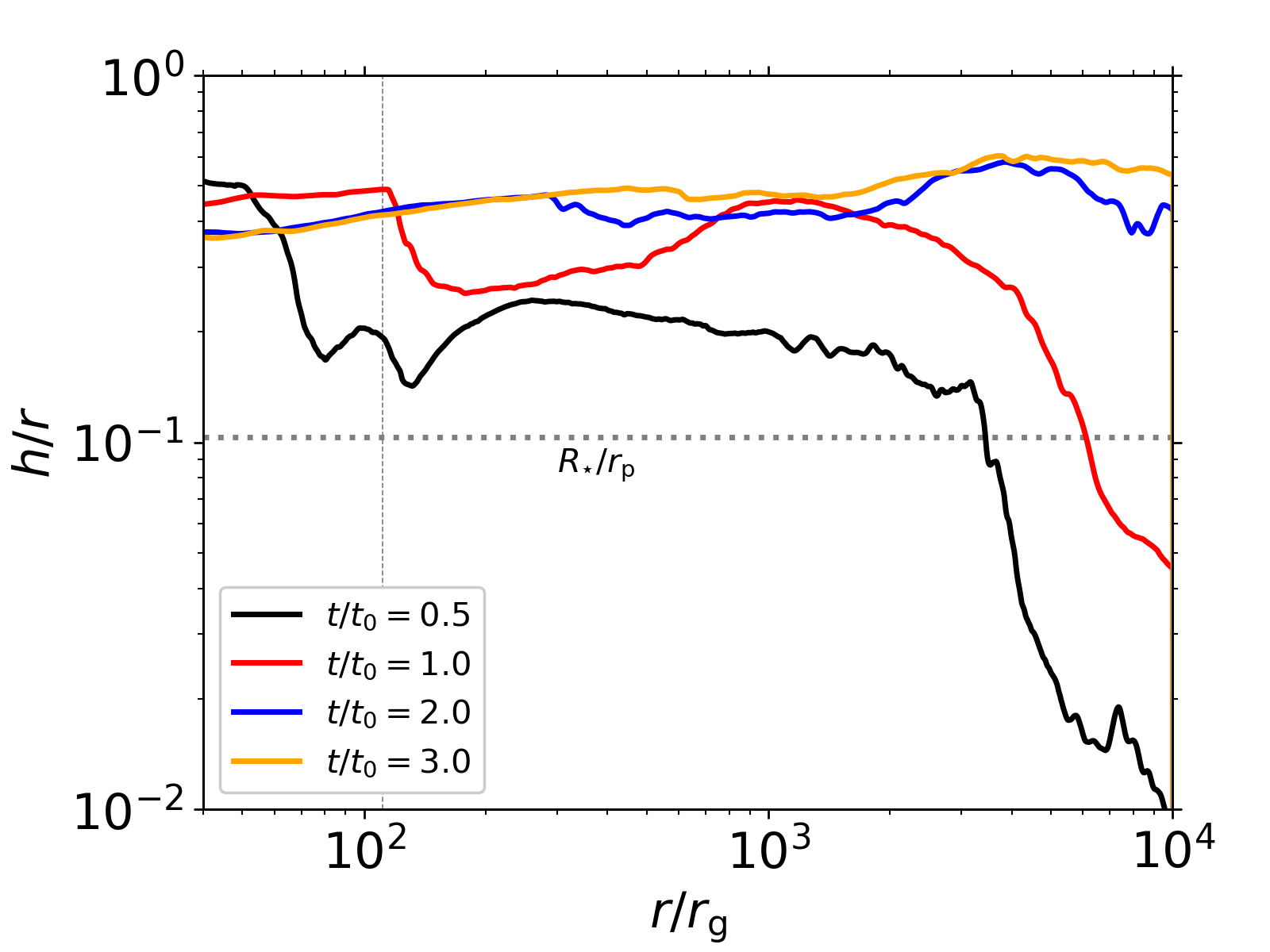}
	\caption{The mass-weighted average of the specific orbital energy of debris (\textit{top-left}), angular momentum (\textit{top-right}), eccentricity (\textit{bottom-left}) and aspect ratio (\textit{bottom-right}) for given $r$ at $t/t_{0}=0.5$, 1, 2 and 3. The dashed grey vertical line in all panels indicate the pericenter distance of the original stellar orbit. The dotted grey horizontal line in some panels show different quantities: $E=\Delta\epsilon$ (\textit{top-left}), initial angular momentum (\textit{top-right}), and $R_{\star}/r_{\rm p}$ (\textit{bottom-right}). The diagonal black line in the \textit{top-left} panel shows a power-law of $r^{-1}$. }
	\label{fig:debris_orbit}
\end{figure*}

\subsection{Formation of shocks}\label{subsub:shocks}

It is convenient to divide the multiple shocks by approximate location: pericenter or apocenter.
The compression and heating of the gas near the pericenter are depicted in detail in Figure~\ref{fig:nozzleshock}. When the incoming stream is narrow and well-defined (the two \textit{upper} panels at $t\lesssim t_{0}$), the nozzle shock structure can be described in terms of two components.  As the different portions of the returning stream converge, adiabatic compression raises the temperature at the center of the stream.  The shock itself runs more or less radially across the stream, extending both inward and outward from the stream center.   However, at later times (beginning at $t \gtrsim t_0$), the structure becomes more complex.  The matter that has been deflected onto lower angular momentum orbits circulates in the region inside $r_{\rm p}$ and develops a pair of nearly stationary spiral shocks.  The shock closest to the position of the nozzle shock stretches progressively farther outward, reaching $\sim 2000r_{\rm g}$ by $t\simeq 2t_0$ (see Figure~\ref{fig:densitydistribution}). However, while the shock extends to greater radii, it also loses strength.  A similar progressive widening and weakening of the nozzle shock were found by \citet{Shiokawa+2015}.

Outgoing previously-returned matter intersects the path of newly-arriving matter in the apocenter region because a combination of apsidal rotation due to the finite duration of the disruption and relativistic apsidal precession cause earlier and later stream orbits to be misaligned \citep{Shiokawa+2015}.   When the apocenter shock first forms, it is relatively close to the black hole ($\gtrsim 1000r_{\rm g}$) because the very first debris to return has orbital energy more negative than $-\Delta E$.   As the mass-return rate rises, its orbital energy also increases, so the debris apocenter moves outward.
However, even at $t\simeq t_{0}$,  when the shock is located at $r\simeq 6000r_{\rm g}$, it is found closer to the BH than the apocenter distance corresponding to $E=-\Delta E$ because the outgoing stream has lost orbital energy to dissipation in the nozzle shock.  At still later times, the apocenter shock moves further inward as the mean energy of the previously shocked matter decreases further.

Some of the outgoing material, upon collision with the incoming stream, is deflected both horizontally and vertically. In the left panel of Figure~\ref{fig:apocentershock}, we show the temperature distribution in the equatorial plane at $t\simeq t_{0}$.  In the \textit{right} panel, one can see a clear boundary between the incoming stream (with temperature $T\simeq 10^{5}$ K) and the outgoing stream (with $T\gtrsim 3 \times 10^{5}$ K).  At this boundary, the outgoing stream is deflected towards the SMBH.

The two panels of Figure~\ref{fig:apocentershock} also reveal that, just as found in \citet{Shiokawa+2015}, the apocenter shock splits into two (called shocks 2a and 2b by Shiokawa et al.).  Shock 2a occurs where the outgoing stream encounters the incoming stream, while shock 2b (visible in these figures as the surface on which the outgoing gas temperature rises from $\simeq 3 \times 10^5$~K to $\simeq 1 \times 10^6$~K) occurs where one portion of the outgoing gas catches up with another portion that has been decelerated by gravity.

\subsection{Debris orbit evolution}\label{subsec:debris}

Undergoing multiple shocks, the debris becomes less eccentric and thicker both in the orbital plane and vertically. To be more quantitative, in Figure~\ref{fig:debris_orbit}, we present the spherically mass-weighted averages of $E/E_{\rm circ}$, angular momentum $l$, eccentricity $e$ and the aspect ratio of debris $h/r$ as functions of $r$ at $t/t_{0}=0.5$, 1, 2 and 3. At $t=0.5t_{0}$, only the debris with $|E|\gtrsim 3\Delta\epsilon$ has returned or is returning, which means only a few $10^{-2}\Msol$ is within $4000 r_{\rm g}$, the apocenter distance for $|E|=3\Delta\epsilon$. At this point, the majority of the mass within $4000 r_{\rm g}$ is unshocked incoming debris; only the earlier-returned, outgoing debris has been shocked once near the pericenter.  Consequently, all the mass-weighted quantities primarily reflect the properties of the newly incoming gas. The debris at $r_{\rm p}\lesssim r < 4000r_{\rm g}$ has $|E| \simeq 3\Delta\epsilon \simeq 0.1E_{\rm circ}$ and angular momentum very close to that of the original stellar orbit, giving an eccentricity almost unity: $1-e\lesssim  0.06$.  Similarly, the aspect ratio of the debris is $\simeq 0.2$, comparable to the value implied by $\Rstar/r_{\rm p}\simeq 0.1$. At $0.5 \lesssim t/t_{0}\lesssim 1$, the shocks begin to dissipate more energy. This results in a small fraction of the gas dropping to orbital energy $\sim -E_{\rm circ}$ and orbiting within $r_{\rm p}$. However, the global orbital properties of the debris ($r\gtrsim r_{\rm p}$) remain largely unchanged. 
By the second half of the simulation, enough orbital energy has been dissipated to reduce the orbital period of much of the gas by a factor order unity (i.e., as previously noted, the dissipated energy becomes comparable to the orbital energy).  The result is a structure close to virialization: the mean orbital energy as a function of radius is $\approx -GM_{\rm BH}/2r$.
 
A similar evolution occurs in the radial distribution of angular momentum $l$.  Until $t \simeq t_0$, the specific angular momentum for nearly all the gas is essentially the same as that of the star.  At later times, however, the shocks have greatly broadened the angular momentum distribution, and the gas has sorted itself with higher angular momentum material located at larger radii.  It is, in fact, this broadening of the angular momentum distribution that leads to the outward radial extension of the nozzle shock noted in the previous subsection.
The combined changes in $E$ and $l$ lead to a decrease in eccentricity from $\gtrsim 0.8 - 0.9$ at $t\simeq t_{0}$ to $0.4-0.5$ at $t \simeq 2 t_0$.  The aspect ratio changes rather little with time over the entire inner region, $r \lesssim 2000 r_{\rm g}$: it rises only from $\simeq 0.2$ to $\simeq 0.5$.  At larger radii, it rises from $\sim 10^{-2}$ immediately after the disruption to $\simeq 0.5$; like the inner region, this is accomplished by $t \simeq 2t_0$.

\begin{figure}
\includegraphics[width=8.6cm]{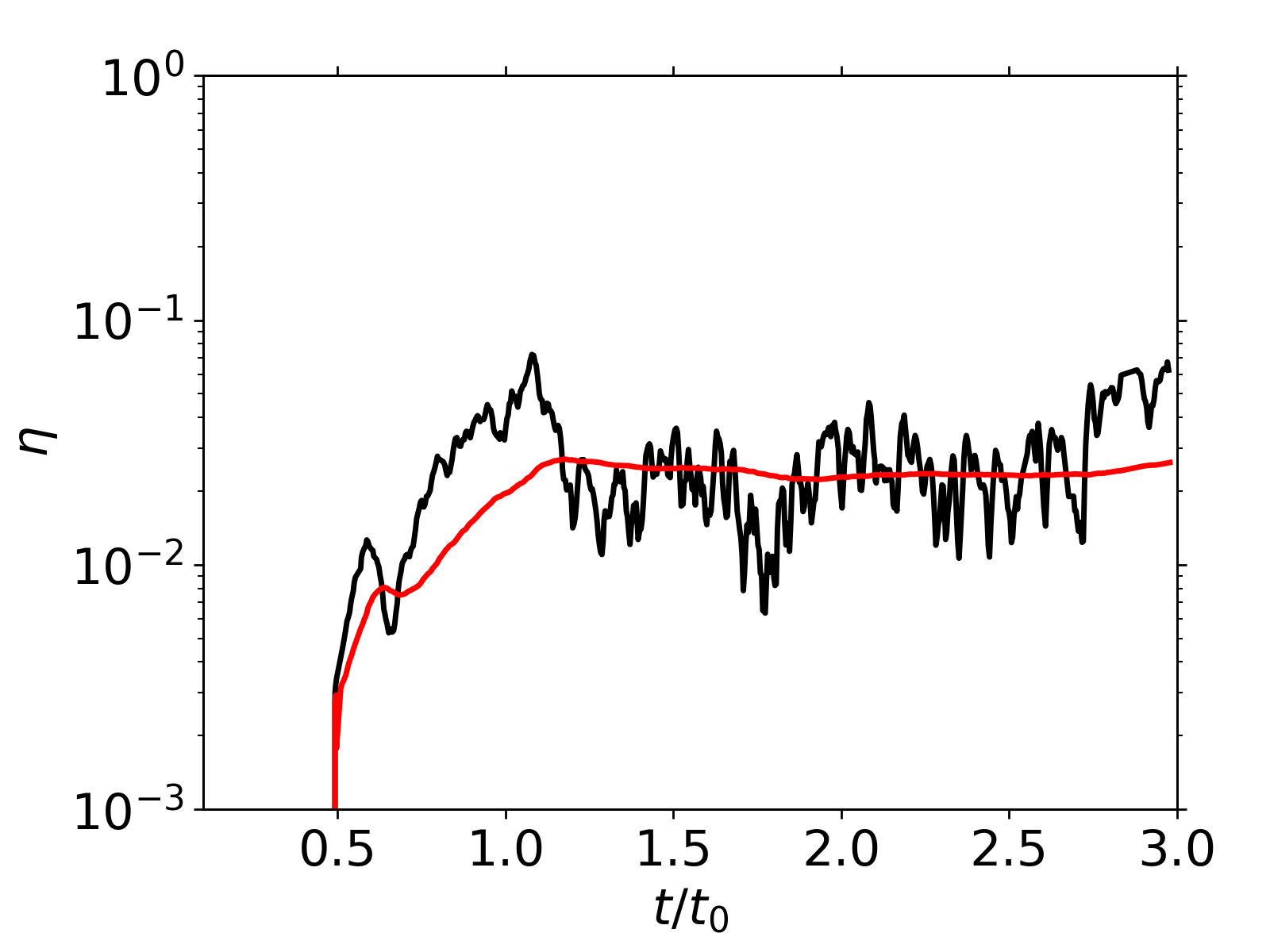}
	\caption{Circularization efficiency $\eta$, defined in Equation~\ref{eq:eta}, as a function of time. The red line shows the time average of $\eta$, estimated as $\eta(t)=\int_{0}^{t} dt^\prime \eta(t^\prime)  / \int_{0}^{t}dt^\prime$.}
	\label{fig:circularization}
\end{figure}

 As mentioned in \S~\ref{sub:overview}, progress toward circularization can be thought of in terms of the rate at which orbital energy is converted into thermal energy.  Evaluating this rate in units of $E_{\rm circ}$ and $t_0$ gives a measure of the circularization ``efficiency":
\begin{align}\label{eq:eta}
\eta \equiv \frac{|dU_{\rm gas+rad}/dt|}{M_{\rm gas} E_{\rm circ}/t_{0}}.
\end{align}
Here $U_{\rm gas+rad}$ is the total thermal (gas + radiation) energy, including thermal energy that has been carried out of the simulation domain by gas flows; from $t \simeq 0.5t_0$ onward, its rate of change is roughly constant at $\simeq 1.4 \times 10^{44}$~erg~s$^{-1}$, integrating to total thermal energy $\simeq 2.2 \times 10^{50}$~erg at the end of the simulation.
$M_{\rm gas}$ is the total gas mass
in the domain at $r<10^{4}r_{\rm g}$. Conveniently, the mass within this region stays very nearly constant over the duration of the simulation (see the \textit{top} panel of Figure~\ref{fig:massenergybudget}), so the total rate of thermal energy creation is very close to directly proportional to $\eta$.  This ``circularization efficiency" as a function of time is shown in Figure~\ref{fig:circularization}\footnote{Note that this definition of $\eta$ differs from the one used by \cite{SteinbergStone2022} who compare the instantaneous energy dissipation rate to the instantaneous fall back rate, $\eta^\prime = |dU_{\rm gas+rad}/dt|(\dot{M}_{\rm fb} E_{\rm circ})^{-1}$. }.
$\eta$ rises very rapidly during the time from $t \simeq 0.5t_0$ to $t\simeq t_{0}$, quickly reaching a maximum $\sim 0.03$. However, from $t \simeq t_0$ to the end of the simulation, its value remains very nearly flat. If it were to stay at that level until circularization was complete, the process would take $\simeq 30 t_{0}$.

\subsection{Radial motion}
\label{subsection:poloidal}

The outer bound of the region in which bound debris is found expands quasi-spherically due to the combined effects of the radiation pressure gradient built by shock heating  and deflection caused by stream-stream collisions. Figure~\ref{fig:verticaldensity} illustrates the azimuthally integrated density at four different times, $t/t_{0}=0.5$, 1, 2 and 3. At $t=0.5 t_0$, the outgoing gas that had been heated by the nozzle shock forms a vertically thick density structure  within $3000r_{\rm g}$. Later on, at $t>0.5t_{0}$, the apocenter shock contributes further  to the expansion of the gas. At $t \gtrsim 2t_{0}$, most of the mass that had been pushed outward starts to fall back towards the SMBH.  

\begin{figure*}\centering
\includegraphics[width=5.7cm]{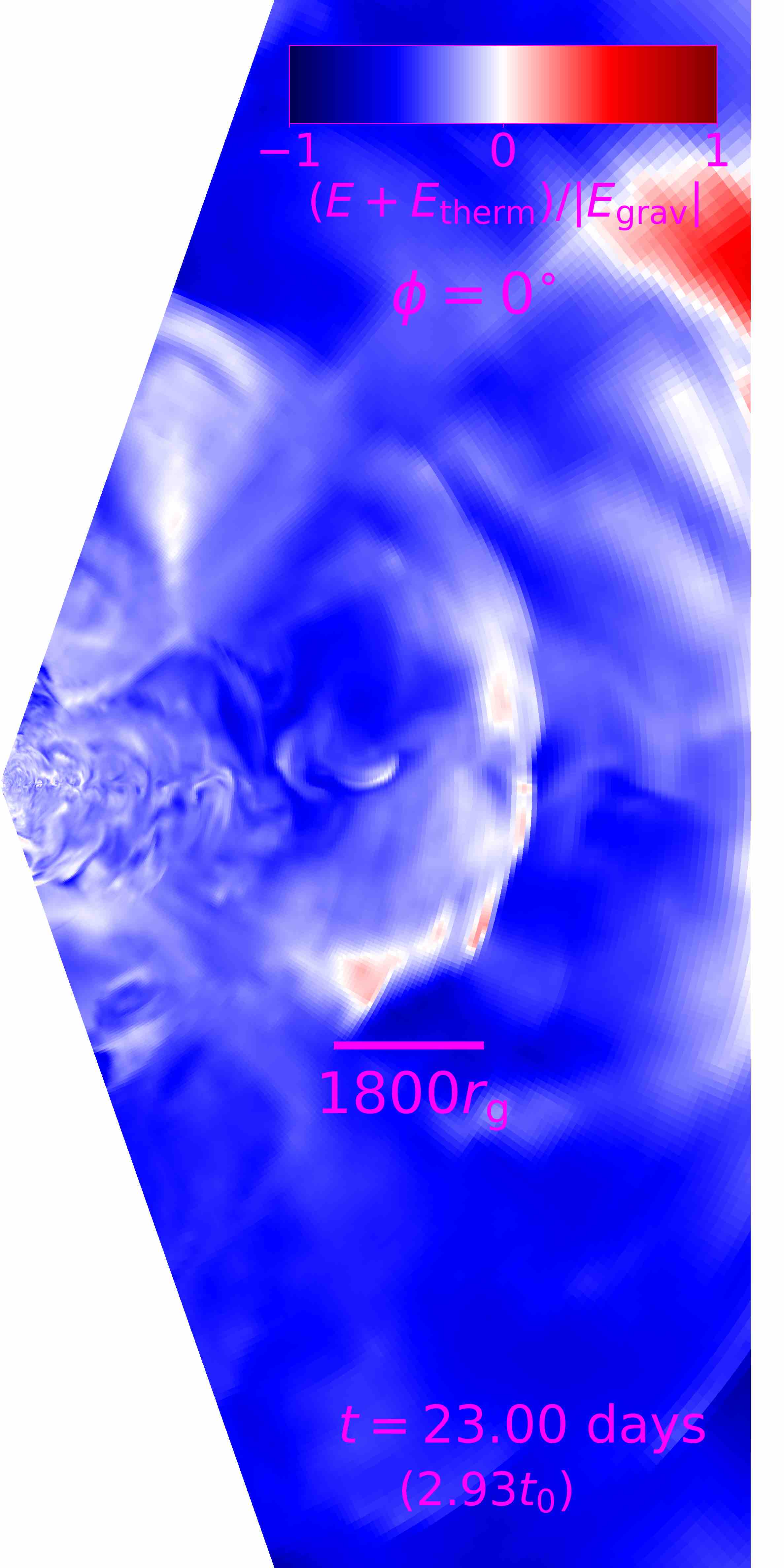}
\includegraphics[width=5.7cm]{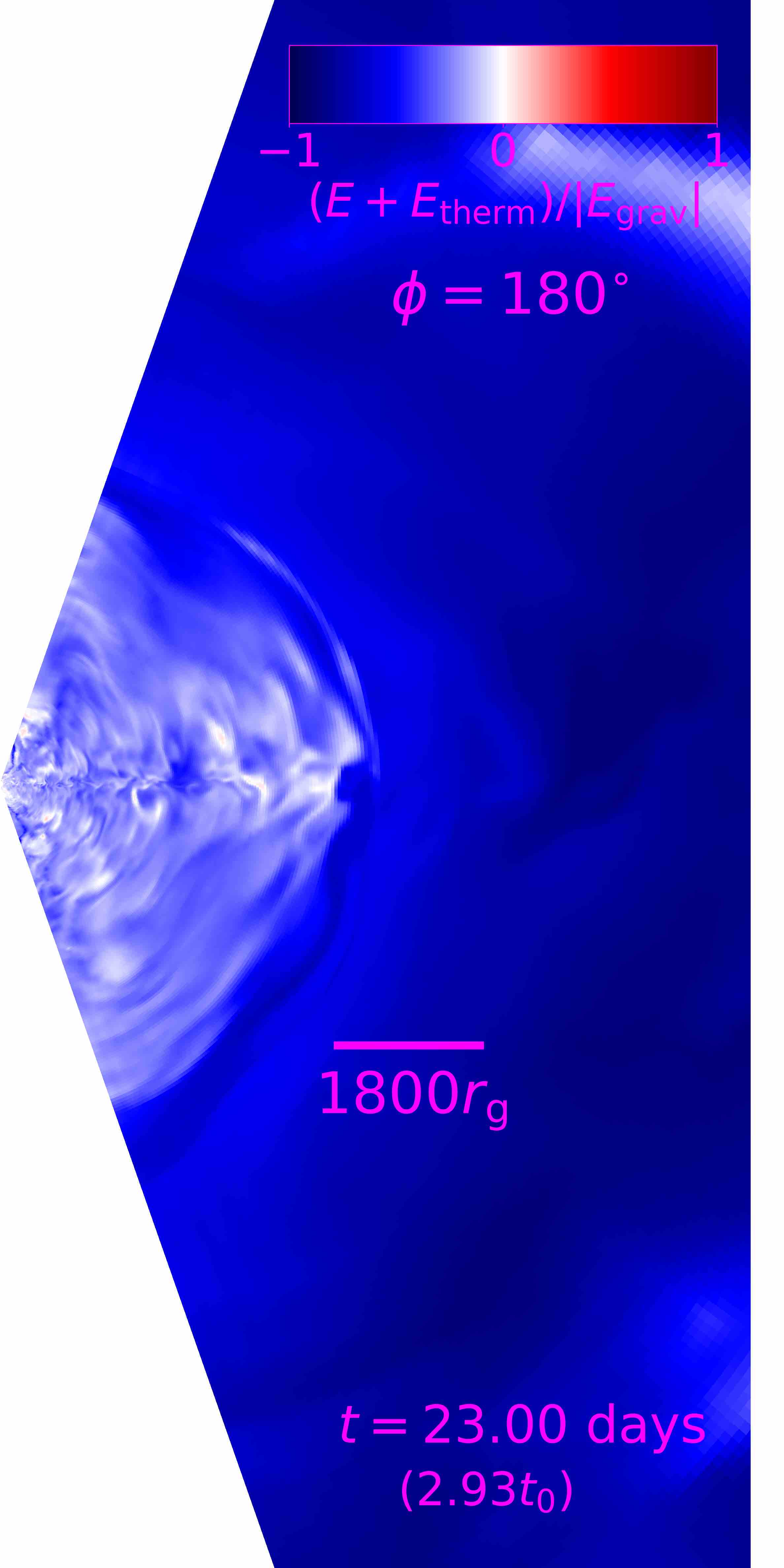}
	\caption{The distribution of the total energy $E+E_{\rm therm}$, normalized by the local gravitational potential $E_{\rm grav}=-G\Mbh/r$, at $\phi=0^{\circ}$ (\textit{left}, near nozzle shock) and $180^{\circ}$ (\textit{right}, near apocenter shock) at $t=3t_{0}$.  }
	\label{fig:verticalenergy}
\end{figure*}

However, in the outermost $\lesssim 1\%$ of the flow, i.e., radii $7000-10000r_{\rm g}$, the gas continues moving outward with a speed of $0.005-0.01c\simeq 1500 - 3000$ km/s even at late times.  To test whether this is an incipient wind, we define unbound gas by the total energy criterion $E+E_{\rm therm} > 0$, where $E$ is the orbital (kinetic and gravitational) energy and $E_{\rm therm}$ is the thermal energy.
\textit{We find almost no matter that has been made unbound after the initial disruption at any time during our simulation (i.e., up to $3t_0$)}. 
 Figure~\ref{fig:verticalenergy} depicts $E+E_{\rm therm}$ at $\phi=0^{\circ}$ (near the nozzle shock) and $180^{\circ}$ (near the apocenter shock) at $t\simeq 3t_{0}$.  As already noted, although essentially all the mass is bound, its specific binding energy is small.  It is therefore not straightforward to predict the final fate of the expanding envelope based on the energy distribution measured at a specific time: energy is readily transferred from one part of the system to another, or from one form of energy to another. However, because the fallback rate declines beyond $3t_0$, the major energy source at later times would be effectively the interactions of gas in the accretion flow that has formed. Hence, the energy distribution in the outer envelope is unlikely to evolve much over time.  In addition, because we allow for no radiative losses, the thermal energy content measured in the simulation data, particularly in the outer layers, is an upper bound to the actual value.  This material's most likely long-term evolution is, therefore, a gradual deceleration followed by an eventual fallback.

\begin{figure}
\includegraphics[width=8.6cm]{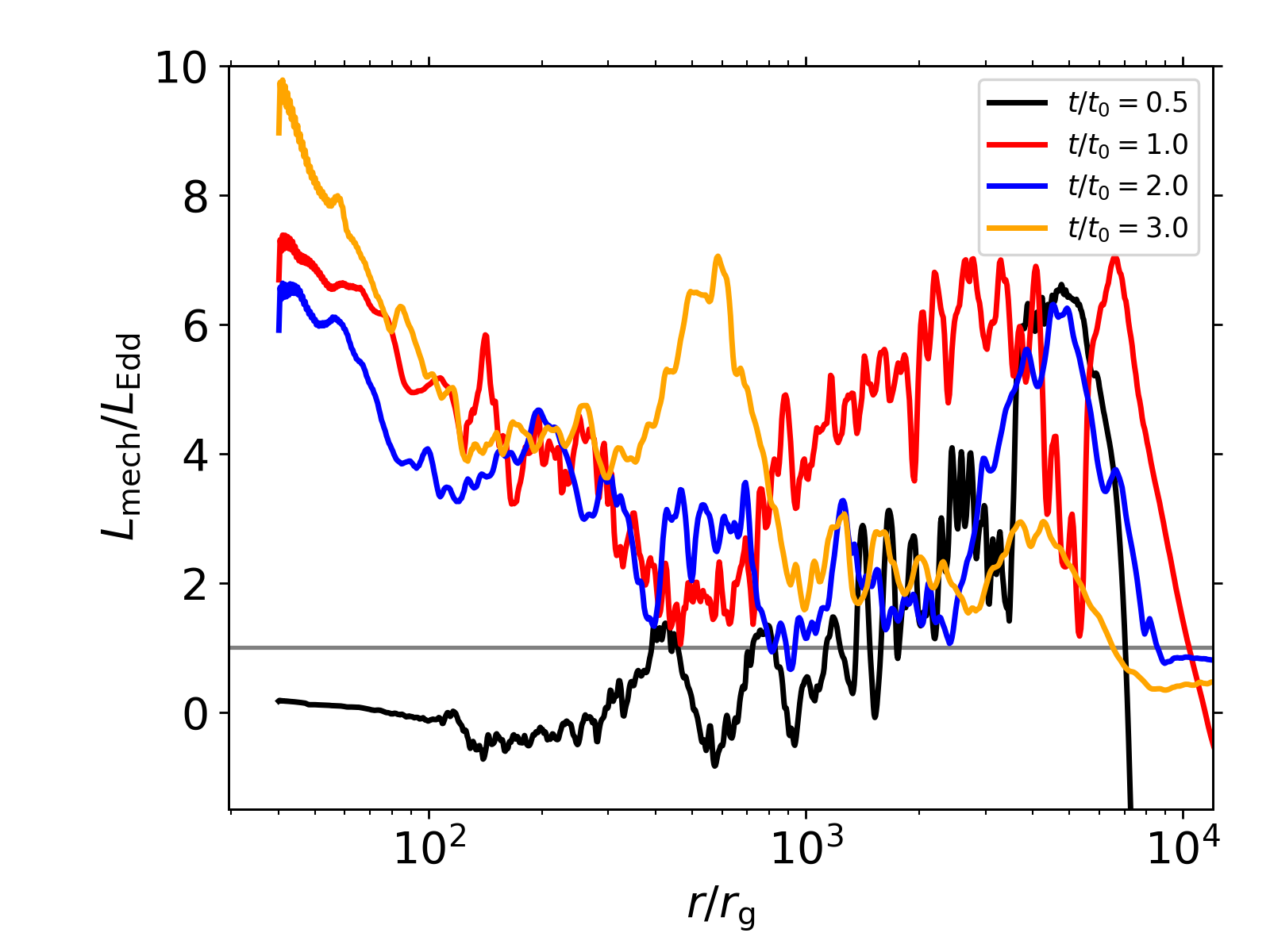}
	\caption{Radial total energy ($E+E_{\rm{therm}}$) flux integrated over a spherical shell at given $r$, normalized by the Eddington luminosity $L_{\rm Edd}$, at four different times. Positive values mean net positive (or negative) energy carried by gas moving radially outward (or inward). }
	\label{fig:lum_mechanical}
\end{figure}

 Moving gas carries  energy.  Defining the mechanical luminosity by
\begin{align}\label{eq:Lmecha}
    L_{\rm mech}(r) = \int_{\Omega} d\Omega r^2 \rho v^{r} (E + E_{\rm therm}), 
\end{align}
we find (as shown in Fig.~\ref{fig:lum_mechanical}) that the net $L_{\rm mech}$ integrated over spherical shells is nearly always positive for $t \gtrsim t_0$ and is super-Eddington. The predominantly negative slope in $L_{\rm mech}(r)$ at $r\lesssim 10^{3}r_{\rm g}$  and $r \gtrsim 5000r_{\rm g}$ indicates that these regions are gaining energy, while the relatively constant mechanical luminosity at $10^{3}r_{\rm g}\lesssim r\lesssim 5 \times 10^3 r_{\rm g}$ shows relatively insignificant energy exchange in that range of radii.

 In interpreting this radial flow of mechanical luminosity, it is important to note that it is due to a mix of outwardly-moving unbound matter and inwardly-moving bound matter.  Both signs of radial velocity are represented on almost every spherical shell; in fact, the mass-weighted mean radial velocity is generally {\it inward} with magnitude $\sim 300 - 1000$~km~s$^{-1}$. Thus, the regions gaining energy do so in large part, but not exclusively, by losing strongly bound mass.

\begin{figure}
\includegraphics[width=8.6cm]{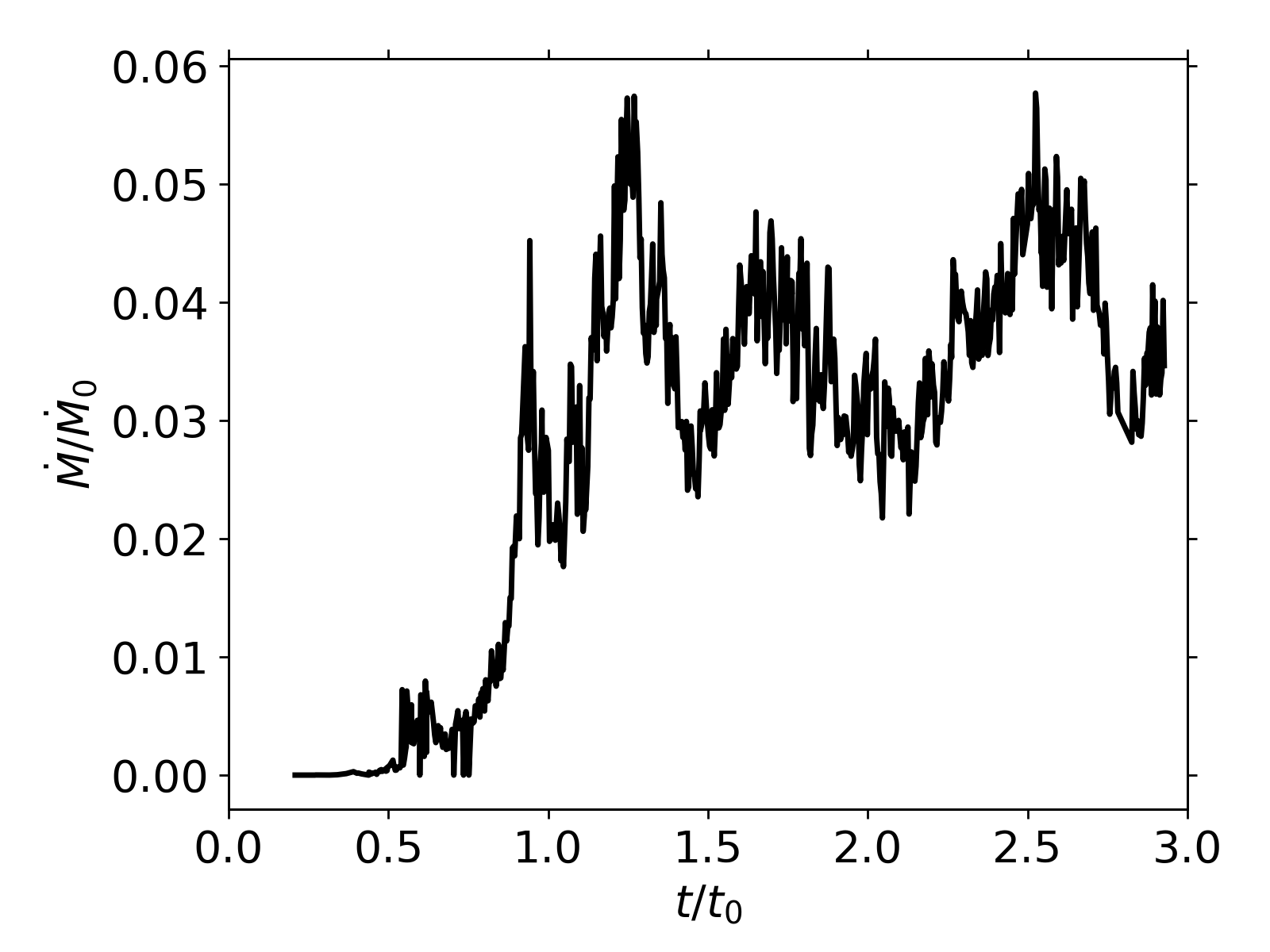}
	\caption{The rate at which gas is expelled through the inner radial boundary at $r=40r_{\rm g}$, divided by $\dot{M}_{0}$, as a function of $t/t_{0}$. }
	\label{fig:massrcpelled}
\end{figure}

\subsection{Matter loss  through inner radial boundary}

Figure~\ref{fig:massrcpelled} shows the rate at which mass falls through the inner radial boundary at $40r_{\rm g}$ and leaves the computational domain. This rate rises rapidly from $t=0$ to $t=t_0$, and from then until $t=3t_0$ fluctuates about a nearly-constant mean value $\simeq 0.035 \dot{M}_0$.
The total lost  mass up to $t\simeq 3 t_{0}$ is $\lesssim 0.025M_{\star}$.  This is a factor $\sim 3$ smaller than the rate, at $t=3t_0$, of mass-loss through a similarly-placed inner cut-out in the simulation of \citet{Shiokawa+2015}.

Although the simulation provides no information about what happens to this gas after it passes within $r=40r_{\rm g}$, we can make certain informed speculations.  By computing its mass-weighted mean specific energy and angular momentum, we find that its mean eccentricity is $\simeq 0.7 - 0.8$ and does not evolve over time.  Its mean pericenter and apocenter distances are $10-15r_{\rm g}$ and $80-120r_{\rm g}$, respectively.  These values imply that the mass lost through the boundary would not accrete onto the SMBH immediately.  In fact, if it follows such an orbit, it should re-emerge from the inner cutout.   If it did so without suffering any dissipation, it would erase the positive energy flux we find at the inner boundary, which is due to bound matter leaving the computational domain.  On the other hand, if it suffered the maximum amount of dissipation consistent with an unchanging angular momentum and settled onto circular orbits inside $40r_g$, it might release as much as $\simeq 10^{51}$~erg, more than enough to unbind the rest of the bound debris, whose binding energy is only $\sim 3 \times 10^{50}$~erg.  However, this estimate of dissipation is an upper bound, and likely a very loose one, with the actual dissipation far smaller.  The location of the shocks suffered by this gas would be much nearer its apocenter at $\sim 100 r_g$ than its pericenter, reducing the kinetic  energy available for dissipation by an order of magnitude or more. 
Oblique shock geometry, as seen in the directly-simulated shocks, sharply diminishes how much kinetic energy is dissipated per shock passage.  In addition, if matter does fall onto weakly eccentric orbits with semimajor axes $< 40 r_{\rm g}$, it will block further inflow, thereby decreasing the net flow across the $r=40r_{\rm g}$ surface.  Whatever accumulation of gas occurs within $\lesssim 100r_g$ is also unlikely to have much effect on the bulk of the bound debris because, as visible in Figure~\ref{fig:nozzleshock}, the radial range at which the bulk of the returning gas passes through the nozzle shock moves steadily outward over time, reaching $\gtrsim 500r_g$ by $t=3t_0$.

\begin{figure*}
\includegraphics[width=9.2cm]{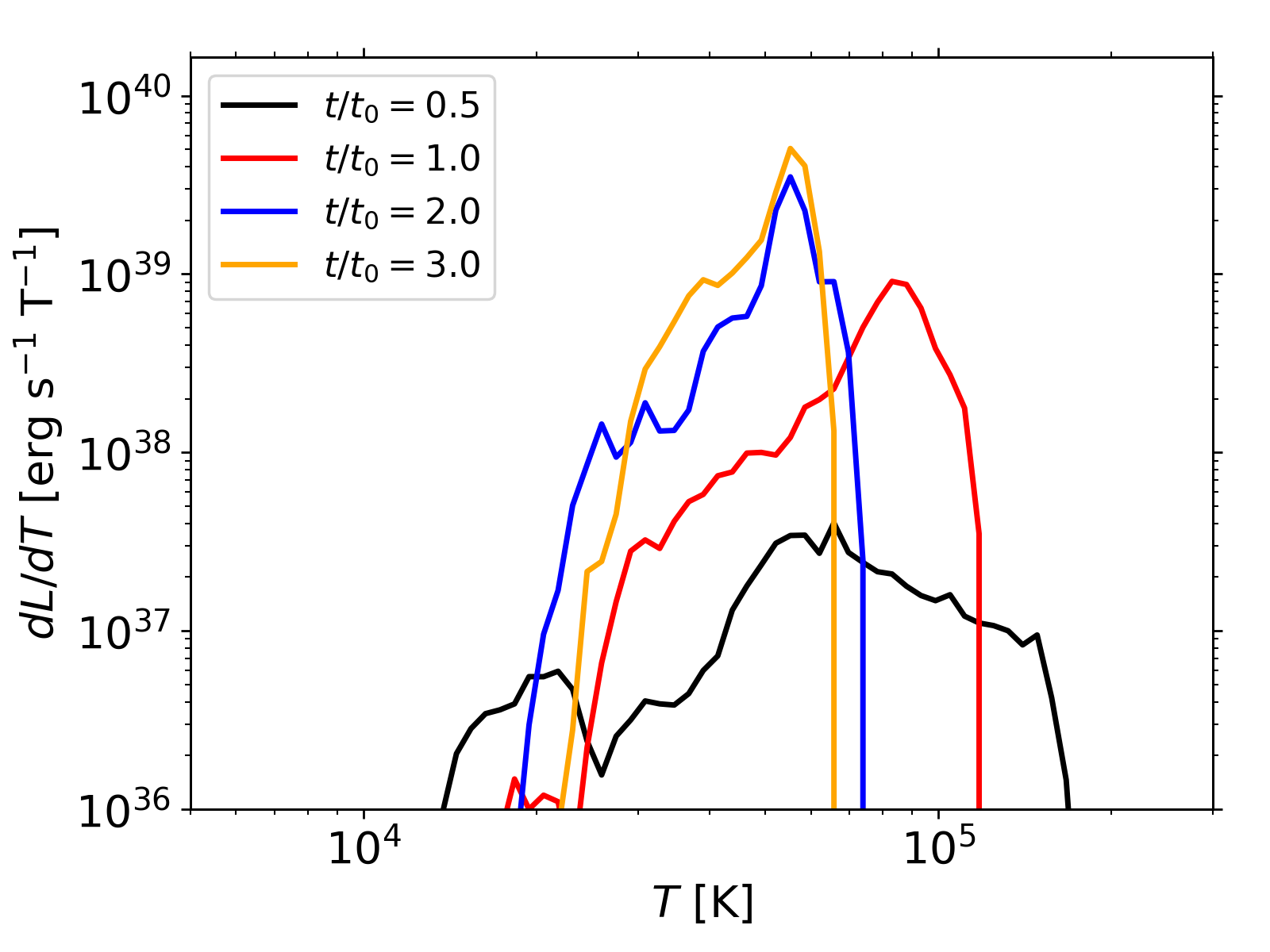}
\includegraphics[width=9.2cm]{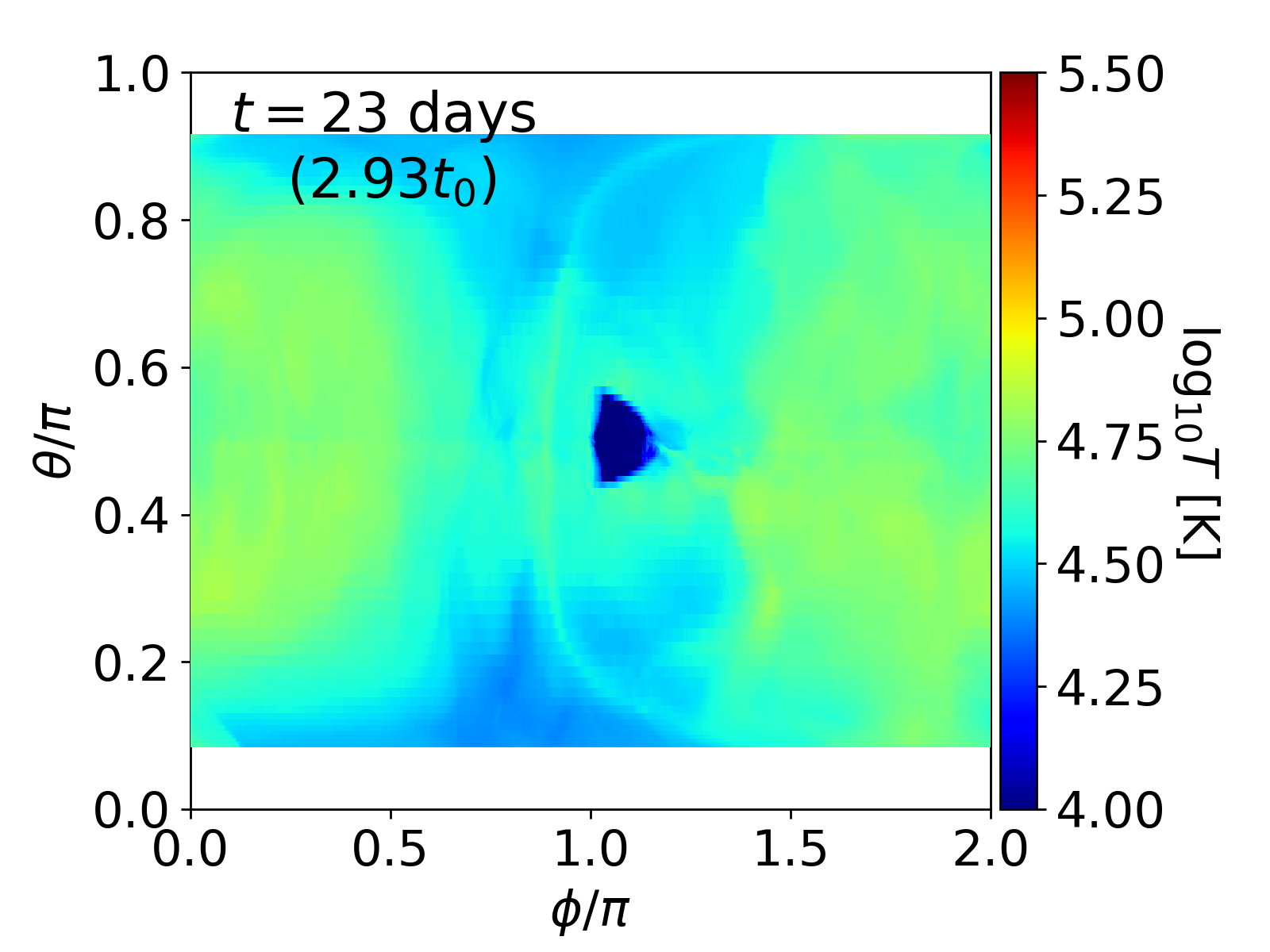}
	\caption{(\textit{Left}) The photosphere temperature distribution $dL/dT$ at four different times $t/t_{0}=0.5$, 1, 2, and 3, and (\textit{right}) the angular distribution of the temperature at $t=3t_{0}$.}
	\label{fig:Teff}
\end{figure*}

\subsection{Radiation}\label{sub:radiation}

To infer the bolometric luminosity of this event, we post-process the simulation results. We first identify the thermalization photosphere with the surface where $\sqrt{\tau_{\rm T}\tau_{\rm ff}}\simeq 1$. Here, $\tau_{\rm T}$ ($\tau_{\rm ff}$) is the Thomson (absorption) optical depth integrated radially inward from the outer $r$ boundary. 
The absorption cross section is calculated using an OPAL table for solar metallicity.
The \textit{upper} panels of Figure~\ref{fig:photosphere_t} show that the photosphere expands quasi-spherically, which is expected from the radial expansion of the outer debris.
At $t=t_{0}$, the photosphere is quasi-spherical located at $r\simeq 4000-5000r_{\rm g}$. It expands to $9000 - 10000 r_{\rm g}$ at $t=2t_{0}$ and  to $\simeq 12000r_{\rm g}$ at $t=3t_{0}$. Given the eccentric orbit of debris, the radius of the photosphere depends on $\phi$ at a quantitative level, even while the its overall shape is qualitatively round.  To demonstrate the $\phi$-dependence, we show in the \textit{bottom} panels of Figure~\ref{fig:photosphere_t} the density distribution and the photosphere at four different azimuthal angles, $\phi=0$, $90^{\circ}$, $180^{\circ}$ and $270^{\circ}$ at $t=3t_{0}$. 

We then estimate the cooling time $t_{\rm cool}$ at all locations inside the photosphere as
  \begin{align}
  t_{\rm cool}(r) = \frac{h_{\rho} \tau(r)}{c } { (1+ u_{\rm gas}/u_{\rm rad})},
   \end{align}
where $h_{\rho}$ is the first-moment density scale height of the gas along a radial path, $\tau$ is the optical depth (radially integrated) to $r$, and  $u_{\rm gas}/u_{\rm rad}$ is the ratio of the local internal energy density to the radiation energy (a ratio that is often only slightly greater than unity).
We then estimate the luminosity by integrating the energy escape rate over the volume within the photosphere, but including only those locations for which $t_{\rm cool}$ is smaller than the elapsed time in the simulation.  This condition accounts for the fact that in order to leave the debris by time $t$, the cooling time from the light's point of origin must be less than $t$.  The resulting expression is
 \begin{align}
     L =\int_{0}^{2\pi}\int_{\theta_{\rm c}}^{\pi - \theta_{\rm c}} \int_{r=r(t_{\rm cool}<t)}^{r=r(\tau=1)}\frac{ a T^{4}}{ t_{\rm cool}} r^{2} \sin\theta dr d\theta d\phi,
 \end{align}
where $a$ is the radiation constant. 

The local effective temperature at each individual cell near the photosphere is then calculated as $T_{\rm ph}=[(dL/dA)/\sigma]^{1/4}$, where $A$ is the surface area of the photosphere and $\sigma$ is the Stefan–Boltzmann constant. We estimate that the peak luminosity is $\simeq 10^{44}$ erg/s $\simeq 10L_{\rm Edd}$, which occurs at $t \simeq t_{0}$.  This is roughly the mean rate of thermal energy creation during the simulation.  The photospheric temperature distribution at $t=0.5t_{0}$ can be described as nearly flat  within the range $\times10^{4}K\lesssim T\lesssim 2\times 10^{5}K$, as shown in the \textit{left} panel of Figure~\ref{fig:Teff}. At $t\gtrsim t_{0}$, the distribution becomes narrower: the distribution at $t\gtrsim 2t_{0}$ has a single peak at $T\simeq (5-6)\times10^{4}$ K. In the right panel of Figure~\ref{fig:Teff} we show the photospheric temperature distribution as a function of observer direction at $t\simeq 3t_{0}$. The temperature is $5-6\times10^{4}$ K over almost the entire photosphere except for a noticeably low-temperature spot at $\phi\simeq\pi$ and $\theta = 0.5\pi$, corresponding to the low-$T$ incoming stream.

\section{Discussion}
\label{sec:discussion}

\subsection{Circularization - fast or slow?}\label{dis:circularization}

The pace of ``circularization" has long played a central role in understanding how TDE flares are powered.  If it is rapid, i.e., takes place over a time $\lesssim t_0$, the debris joins a small ($r \lesssim r_{\rm p}$) accretion disk as soon as it first returns.  In addition, accretion takes place on a timescale short compared to $t_0$ because the orbital period on this scale is shorter than $t_0$ by a factor $(\Mstar/\Mbh)^{1/2} \sim 10^{-3}$.  Even after waiting $\sim 10$ orbital periods for MRI turbulence to saturate and then consuming many more orbital periods to flow inward by magnetic stresses, the total inflow is still short compared to $t_0$.   The dissipation rate at the time of peak mass-return would then be strongly super-Eddington.

The result of our simulation, however, is that ``circularization" is actually very slow.  We find that the returning debris forms a large cloud that stretches all the way from the pericenter of the original stellar orbit to the apocenter of the most-bound debris, a dynamic range $\sim 30 - 100$.   Throughout the first $3t_0$ after disruption, only a small fraction of the debris resides within the pericenter. 
The mass-weighted mean eccentricity falls from $\simeq 0.9$ to $\simeq 0.4 - 0.6$ by $t\gtrsim 2t_{0}$, but doesn't decrease further from that time to until at least $3t_0$.
Thus, by this late time the debris has neither achieved a circular orbit nor been compressed within $r \sim r_{\rm p}$.

Such slow circularization is consistent with the low energy dissipation rate. The thermal energy within the system is very small compared with $E_{\mathrm{circ}}$, the energy that must be removed from the bound debris' orbital energy in order to fully circularize it. Similarly, the circularization efficiency parameter $\eta$ suggests that several dozen $t_0$ are required in order to dissipate $E_{\mathrm{circ}}$ of energy (see Figure \ref{fig:circularization}).  Thus, we may conclude that little circularization is accomplished during the time in which most of the debris mass returns to the BH.

Our conclusions in this matter agree with earlier findings of \cite{Shiokawa+2015}, who used a somewhat cruder computational scheme and less realistic conditions (these authors considered a disruption of a white dwarf of $0.64 \Msol$ by a BH of $500 \Msol$).
On the other hand, they differ with those  of \cite{SteinbergStone2022}, who analyzed the ``circularization efficiency" in terms of the heating rate per returning mass rather than our definition, the heating rate per {\it returned} mass over a time $t_0$.  On the basis of tracking this definition of circularization efficiency up to $t = t_0$, they argued that it was growing exponentially on a timescale $\sim t_0$, so that full circularization might be achieved quickly.
Interestingly, our definition of efficiency also grows rapidly with time during the first $t_0$; in this respect we agree with \citet{SteinbergStone2022}.  However, we also find that it flattens out shortly after $t_0$.  Thus, one possible explanation of the contrast in our conclusions about the magnitude of energy dissipation is simply that our simulation ran longer than theirs when measured in $t_0$ units.  It is also possible that some of the difference in the results could be attributed to differences in our physical assumptions. 
\citet{SteinbergStone2022} used a spherical harmonic oscillator potential at $r < 30~r_{\rm g}$ (private conversation with Elad Steinberg) and a Paczynski-Wiita potential at larger radii, whereas we used a Schwarzschild spacetime with a cut-out at $40r_{\rm g}$; they described radiation transport by a flux-limited diffusion scheme, whereas we included radiation only as a contribution (often the dominant one) to the pressure.  On balance, though, because the gravity descriptions used are not very different on the relevant lengthscales and the long cooling times in the system severely limit radiative diffusion, these contrasts are unlikely to explain this disagreement.   Lastly, it is possible that the difference in parameters (our $M_{\rm BH} = 10^5 M_\odot$ and $M_* = 3M_\odot$ vs. their $M_{\rm BH} = 10^6 M_\odot$ and $M_* =1 M_\odot$) may also play a role.  Further simulations will be necessary in order to test this possibility.

\subsection{Energy Dissipation: Shocks vs. Accretion}

The physical assumptions in our simulation restrict the creation of thermal energy to two mechanisms: shocks and compressive work done within the fluid.  There is no energy release due to classical accretion because our equations contain neither MHD turbulence nor phenomenological viscosity.   Nonetheless, we have demonstrated that shocks and compression can, without these other processes, generate enough energy during a few $t_0$ to power the observed luminosity of TDEs.   We estimate a photon luminosity during this period of $\sim 10^{44}$~erg~s$^{-1}$, and all of this energy was generated by shocks and compressive work. As discussed in the previous subsection, we have demonstrated the {\it absence} of orbital energy loss that is a prerequisite for forming a classical accretion flow.

\subsection{Outflow}

A third interesting finding is that we do not find a significant unbound outflow emerging from the bound debris.  Very nearly all the bound material that has returned to the vicinity of the SMBH remains bound by the end of our simulations.  Although we do see outward motion, its slow speed indicates that the material remains bound (see Figure~\ref{fig:verticalenergy}). It should therefore eventually slow down and fall back.

This result places an even stronger upper bound on the dissipated energy than the earlier result that there was too little dissipation to circularize the matter, as the specific energy needed to unbind the debris is significantly smaller than that needed to circularize it around the original pericenter.  Whereas the circularization energy is $\sim 3 \times 10^{51}$~erg, the binding energy is only $\sim 3 \times 10^{50}$~erg.  
That almost no initially bound debris is rendered unbound is consistent with observational limits on outflows from both radio and optical TDEs \citep{MatsumotoPiran2021}. 

This conclusion, which is contrary to a number of predictions \citep[e.g.,][]{Jiang+2016,Bonnerot+2021,Huang+2023},
also casts some doubt on the possibility \citep{MetzgerStone2016} that the kinetic energy of an outflow is the solution to the ``inverse energy crisis" mentioned earlier.  When the source of heating is shocks, we find negligible transport of energy to infinity associated with outflows.  Interestingly, although \citet{SteinbergStone2022} do find an unbound outflow, its mechanical luminosity is only $\simeq 1.6 \times 10^{42}$~erg~s$^{-1}$ if the outflow velocity they quote, 7500~km~s$^{-1}$, is its velocity at infinity.  This is such a small fraction of the heating rate that even this sort of wind does not play a significant role in the energy budget.  Moreover, even if {\it all} the mass lost through our inner boundary were quickly accreted in a radiatively efficient manner, as we have already estimated, the associated heat produced would be only a factor of 4 -- 5 greater than the thermal energy generated by  shocks in the first $3t_0$ after the disruption.  In this sense, we have also placed a strong limit on the ability of a wind dependent upon accretion  energy to carry away a large  quantity  of energy.

\subsection{The ultimate fate of the bound debris}

Our simulation ends at $3t_0$ with nearly all the bound debris $\sim 10^3 - 10^4 r_{\rm g}$ from the BH, spread over a large eccentric cloud.  The question naturally arises: what happens next?  Extrapolating from their qualitatively similar results, \citet{Shiokawa+2015} suggested that, after the usual $\sim 10$ orbital-period time necessary for saturation of MHD turbulence driven by the magnetorotational instability (MRI), the gas would accrete in more or less the fashion of circular accretion disks.

Since that work, it has been shown \citep{Chan+2018,Chan+2022} that, indeed, the MRI is a genuine exponentially-growing instability in eccentric disks and, in its nonlinear development, creates internal magnetic stresses comparable to those seen in circular disks.   However, its outward transport of angular momentum may, in the context of eccentric disks, cause the innermost matter to grow in eccentricity while outer matter, the recipient of the angular momentum removed from the inner matter, becomes more circular \citep{Chan+2022}.

If this is the generic result of MRI-driven turbulent stresses in an eccentric disk, accretion might be radiatively inefficient, as matter can plunge directly into the BH if it has sufficiently small angular momentum \citep{Svirski+2017}.  The condition for this to happen is for the angular momentum transport to be accompanied by very little orbital energy loss.  It is then possible for fluid elements of very low angular momentum to fall ballistically into the SMBH after having radiated only a small amount of energy. In this case, the system will dim rapidly after the thermal energy created by shocks has diffused out in radiation. 

However, it remains to be determined whether this is, in fact, the situation in TDE eccentric accretion flows.   If, instead, the work done by torques associated with angular momentum transport is substantial, a compact, more nearly circular, accretion disk eventually forms.  This disk will then behave much more like a conventional accretion flow, radiating soft X-rays until most of the disk mass has been consumed. 

If the energy lost per unit accreted mass comes anywhere near the $\sim 0.1c^2$ of radiatively-efficient accretion, the total energy radiated over this prolonged accretion phase could be quite large:  $0.1 M_{\star} c^2  \approx 10^{53}$~erg.  However, sufficiently long accretion timescales might keep the luminosity relatively low.  There is some observational evidence for such radiation on multi-year timescales, both in X-rays \citep[e.g.,][]{Jonker+2020, Kajava+2020}  and UV \citep[e.g.,][]{vanVelzen+2021,Hammerstein+2023}. In these long-term observations, the luminosity declines gradually enough ($\propto t^{-1}$) to make the total energy radiated logarithmically divergent.

A related question is posed by the matter that passed through our inner radial boundary.  To the extent that some portion of it does dissipate enough energy to achieve a near-ISCO orbit, there is the possibility of significant energy release in excess of what was seen in our simulation.  In fact, in order to generate soft X-ray luminosities comparable to those often seen ($\sim 10^{44}$~erg~s$^{-1}$ at peak), all that is required is a mass accretion rate $\sim 3 \times 10^{-3}\Mstar/t_0$.  Thus, if $\sim 0.3$ of the matter passing through our inner boundary were able to accrete onto the BH, it might be able to account for the X-ray luminosity sometimes seen, given an optically thin path to infinity.  For the parameters of our simulation, there appears to be little or no solid angle through which such a path exists (see Figure~\ref{fig:photosphere_t}), but, as shown by \citet{Ryu+2020e}, the ratio $t_{\rm cool}/t_0$ falls to $\lesssim O(1)$ when $\Mbh \gtrsim 10^6 M_\odot$.  Consequently, radiative cooling might make the flow geometrically thinner for larger $\Mbh$ events, permitting X-rays emitted near the center to emerge during the time of the optical/UV flare.  Alternatively, for those cases that, like our simulation, have relatively long cooling times, X-ray emission may become visible only after a significant delay relative to the optical/UV light, a delay that has been observed in several TDEs \citep{Gezari+2017,Kajava+2020,Hinkle+2021,Goodwin+2022}.

\subsection{Comparison with \citet{Ryu+2020e}}

\citet{Ryu+2020e} introduced a parameter-inference method {\sc TDEmass} for $\Mbh$ and $\Mstar$ built on the assumption that optical TDEs are powered by the energy dissipated by the apocenter shock. In this method, one assumes that the peak luminosity and temperature occur at $t\simeq 1.5t_{0}$ when the most-bound debris collide with the incoming stream at the apocenter.
Using our numerical results to determine the two parameters of {\sc TDEmass} (setting $c_1$, the ratio of the photospheric radius to the apocenter distance, to 1.2 and the solid angle of the photosphere to $4\pi$),
we find that the luminosity and temperature at the peak of the bolometric lightcurve would be $3\times10^{44}$ erg/s and 70000 K \citep[see Equations. 1, 2, 6 and 9 of][]{Ryu+2020e}.

These values can be compared with the estimates derived from our cooling time method, $L \approx 10^{44}$~erg~s$^{-1}$ and $T\approx 60000$~K, measured at $t\simeq 1.5t_{0}$.
The contrast in luminosity may be a consequence of an assumption made in the method of \citet{Ryu+2020e}: that the heating due to shocks is radiated promptly. Although this is a reasonable approximation for $\Mbh \gtrsim 10^6 M_\odot$,  our simulation has shown that when $\Mbh$ is as small as $\sim 10^5M_\odot$, cooling is significantly retarded (in fact, \citealt{Ryu+2020e} pointed out that $t_{\rm cool}/t_0 \propto \Xi^{5/2} \Mbh^{-7/6} \Mstar^{4/9}$).  
Although our simulation suggests that this method may require some refinement in the range of small SMBH masses, overall, whether with or without the corrections suggested by the detailed numerical simulation,  the peak luminosity is in the range of {optical/UV bright} TDEs. The temperature estimated from the simulation is larger by only a factor of 1.2, which is reasonable given the approximate treatment of the radiation in our scheme.

\section{Conclusions}
\label{sec:Summary}

Following the energy often provides a well-marked path toward understanding the major elements of a physical event.   It is especially useful for TDEs because one might define their central question as ``How does matter whose initial specific orbital energy is $\sim 10^{-4}c^2$ dissipate enough energy to both power the observed radiation and then, in the long-run, fall into the black hole?"

This question can be made more specific by pointing out certain milestones in energy.  In a typical TDE flare, $\sim 3 \times 10^{50}$~erg is radiated during its brightest period, although in a number of cases an order of magnitude more is radiated over multiple year timescales \citep[e.g.,][]{vanVelzen+2021,Hammerstein+2023}.  The immediately post-disruption binding energy of the bound gas in the simulation described here is very similar to this number, $2 \times 10^{50}$~erg.   The energy required to circularize all the bound gas is 1.5~dex larger, $7.5 \times 10^{51}$~erg.  Lastly, the energy that might be liberated through conventional relativistic accretion of all the bound material is $\sim 3 \times 10^{53}$~erg.

Comparing the results of our simulation---$\sim 1.5 \times 10^{50}$~erg radiated over a time $3t_0$ long and final gas binding energy less than a factor of 2 greater than in the initial state ($3 \times 10^{50}$~erg)---to these milestones points to a number of strong implications.

First, and most importantly, the radiation we estimate as arising from our simulation is very close to the typical radiated energy during the brightest portion of the flare.   In other words, the hydrodynamics we have computed, in which shocks dissipate orbital energy into heat, succeed in matching the most important quantity describing TDE flares.

Second, over this period the binding energy of the debris does not change appreciably.   It immediately follows from the virial theorem that the scale of the region occupied by the debris likewise does not change appreciably.   The only modification that might be made to this conclusion is that radiation losses would increase the binding energy by a factor of $\sim 1.5 - 2$.   The area of the photosphere is determined by the scale of the region containing the bound debris.

Third, swift ``circularization", that is, confinement of the bound debris to a circular disk with outer radius $\sim r_{\rm p}$, does not happen.  This process requires the bulk of the debris to increase its binding energy by a factor $\sim 30$; this did not happen.

Fourth, radiatively efficient accretion of most of the debris mass onto the black hole certainly did not happen.  If this had occurred, the mass remaining on the grid would be substantially smaller, and the energy released would have rendered the remaining mass strongly unbound, as it corresponds to a total dissipated energy $\sim 10^3\times$ larger than seen.

Lastly, we have also found that, contrary to some expectations, essentially no debris gas that was bound immediately after the disruption was rendered unbound by shock dynamics.

\section*{Acknowledgements}

We are thankful to the anonymous referee for constructive comments and suggestions. We thank Elad Steinberg and Nick Stone for helpful conversations.  We also thank Suvi Gezari for informing us about delayed X-ray flares in TDEs.
This research project was conducted using computational resources (and/or scientific computing services) at both the Texas Advanced Computing Center and the Max-Planck Computing \& Data Facility. At TACC, we used Frontera under allocations PHY-20010 and AST-20021.  In Germany, the simulations were performed on the national supercomputer Hawk at the High Performance Computing Center Stuttgart (HLRS) under the grant number 44232. TP is supported by ERC grant MultiJets.  JK is partially supported by NSF grants AST-2009260 and PHY-2110339.

\software{
python \citep{python3}; matplotlib \citep{Hunter:2007}; \mesa~\citep{Paxton+2011}; 
\harm ~\citep{Noble+2009}.
}

\end{document}